\documentclass[acmsmall]{acmart}

\usepackage{algorithm}
\usepackage{algpseudocode}
\usepackage{graphicx}
\usepackage{textcomp}
\usepackage{xcolor}
\usepackage{booktabs} 
\usepackage{graphicx}
\usepackage{subfigure}
\usepackage{multirow}
\usepackage{bm}
\usepackage{braket}
\usepackage{mathtools}
\usepackage{enumerate}
\usepackage{enumitem}
\usepackage{empheq}
\usepackage{calc}
\usepackage{balance}
\usepackage{dsfont}
\usepackage{textcomp} 
\usepackage{caption}
\usepackage{color, colortbl}
\usepackage{adjustbox}
\usepackage{url}
\usepackage[utf8]{inputenc}
\usepackage[title]{appendix}

\AtBeginDocument{%
  \providecommand\BibTeX{{%
    \normalfont B\kern-0.5em{\scshape i\kern-0.25em b}\kern-0.8em\TeX}}}

\setcopyright{acmcopyright}
\copyrightyear{2023}
\acmYear{2023}
\acmDOI{XXXXXXX.XXXXXXX}
\acmJournal{JACM}
\acmVolume{00}
\acmNumber{0}
\acmArticle{000}
\acmMonth{0}

\begin{document}

\title{Reconstructing Turbulent Flows Using Physics-Aware Spatio-Temporal Dynamics and Test-Time Refinement}

\author{Shengyu Chen}
\affiliation{%
  \institution{University of Pittsburgh}
      \city{Pittsburgh}
  \country{United States}
}
\author{Tianshu Bao}
\affiliation{%
  \institution{Vanderbilt University}
      \city{Nashville}
  \country{United States}}
\author{Peyman Givi}
\affiliation{%
  \institution{University of Pittsburgh}
      \city{Pittsburgh}
  \country{United States}}
\author{Can Zheng}
\affiliation{%
  \institution{University of Pittsburgh}
      \city{Pittsburgh}
  \country{United States}}
\author{Xiaowei Jia}
\affiliation{%
  \institution{University of Pittsburgh}
    \city{Pittsburgh}
  \country{United States}}

\renewcommand{\shortauthors}{}

\begin{abstract}
Accurate simulation of turbulent flows is of crucial importance in many branches of science and engineering.  Direct numerical simulation (DNS) provides the highest fidelity means of capturing all intricate physics of turbulent transport.  However, the method is computationally expensive because of the  wide range of turbulence scales that must be accounted for in such simulations.    Large eddy simulation (LES) provides an alternative. In such simulations,  the large scales of the flow are resolved and the effects of small scales are modelled.    Reconstruction of the DNS field from the low-resolution LES is needed for a wide variety of applications.  Thus the construction of super-resolution (SR) methodologies that can provide this reconstruction has become an area of active research.  In this work, a  new physics-guided neural network is developed for such a reconstruction. The method leverages the partial differential equation that underlies the flow dynamics in the design of spatio-temporal model architecture. A degradation-based refinement method is also developed to enforce physical constraints and to further reduce the accumulated reconstruction errors over long periods. Detailed DNS data on two turbulent flow configurations are used to assess the performance of the model.  
\end{abstract}





\maketitle

\section{Introduction}

Direct numerical simulation (DNS) of the Navier-Stokes equations is a brute-force computational method and is  the method with the highest reliability for capturing turbulence dynamics~\cite{Givi94}. The computational cost of such simulations is very expensive for flows with high Reynolds numbers. Large eddy simulation (LES) is a popular alternative, concentrating on the larger scale energy-containing eddies, and filtering the small scales of transport~\cite{Sagaut05}. In this way,  LES can be conducted on coarser grids as compared to  DNS, but obviously with  less fidelity~\cite{NNGLP17}. 

Machine learning, including super-resolution (SR) methods~\cite{Cheo2003SR}, have been advocated as a means of reconstructing highly resolved DNS from LES data. These methods have shown tremendous success in reconstructing high-resolution data in various commercial applications. The majority of current SR models use convolutional network layers (CNNs)~ \cite{albawi2017understanding} to extract representative spatial features and transform them through complex non-linear mappings to recover high-resolution images. Starting from the end-to-end convolutional SRCNN model~\cite{dong2014learning}, several investigators  have explored the addition of other structural components such as skip-connections~\cite{zhang2018image, zhang2018residual, ahn2018fast, Duong2021, Dai2019, Duong2021}, channel attention~\cite{zhang2018image}, adding adversarial training objectives~\cite{ledig2017photo, chen2017fsrnet, wang2018recovering, wang2018esrgan, karras2018progressive, gan8759375, cheng2021mfagan, Long2021}, and more recently, Transformer~\cite{parmar2018image}-based SR methods
~\cite{fang2022cross,yang2020learning, lu2022transformer, fang2022hybrid, wang2022detail, zou2022self, liang2022light}.

Given their success in computer vision, SR methods are becoming increasingly popular in turbulence  reconstruction~\cite{liu2020deep,xie2018tempogan, Deng2019SuperresolutionRO, wang2018esrgan, fukami2019super, Fukami_2020}. Despite their popularity, these methods face some limitations when it comes to representing continuous flow dynamics in the spatial and temporal fields using discrete data samples. Consequently, they can learn spurious patterns between sparse observations, which often lack generalizability. Additionally, the training of SR models is hindered by the scarcity of high-fidelity DNS data due to the required high computational cost of such simulations.

In this work, a novel method termed the ``continuous networks using differential equations'' (CNDE) is developed to improve the SR reconstruction. This development is by leveraging the underlying physical relationships to guide the learning of generalizable spatial and temporal patterns in the reconstruction process. The method consists of three components: the Runge-Kutta transition unit (RKTU), the temporally-enhancing layer (TEL), and degradation-based refinement. The RKTU structure is designed based on the governing partial differential equations (PDEs) and is used for capturing continuous spatial and temporal dynamics of turbulent flows. The TEL structure is designed based on the long-short-term memory (LSTM)~\cite{LSTM} model and is responsible for capturing long-term temporal dependencies.  The degradation-based refinement is to adjust the reconstructed data by enforcing consistency with physical constraints. 

Model appraisal is made by considering detailed data sets pertaining to two turbulent flow configurations: (1) a forced isotropic turbulent (FIT) flow~\cite{FITdata}, and (2) the Taylor-Green vortex (TGV) flow~\cite{brachet1984taylor}.  The results of the consistency assessments demonstrate the capability of the CNDE in terms of the reconstruction performance over space and time.  The effectiveness of each component of the methodology is demonstrated qualitatively and quantitatively.

\section{Related Work}
\subsection{Super-Resolution}

Single image super-resolution (SISR) via deep learning has been the subject of many investigations in computer vision. These methods derive their power primarily from the utilization of convolutional network layers~\cite{albawi2017understanding}, which extract spatial texture features and transform them through complex non-linear mappings to recover high-resolution data. One of the earliest SR methods for SISR is SRCNN~\cite{dong2014learning}, which learns an end-to-end mapping between coarse-resolution and high-resolution images by employing a series of convolutional layers. Another scheme is the skip-connection layers~\cite{Duong2021, Dai2019, zhang2018residual, ahn2018fast, Tai2017},  which enable the bypassing of abundant low-frequency information and emphasize the relevant information to improve the stability of the optimization process in deep neural networks. 
Several investigators have explored the adversarial training objective by using the generative adversarial network (GAN) for  SISR. For example, the SRGAN model~\cite{ledig2017photo} stacks the deep residual network to build a deeper generative network for image super-resolution and also introduces a discriminator network to distinguish between reconstructed images and real images using an adversarial loss function. The ultimate goal is to train the generative network in a way that the reconstructed images cannot be easily distinguished by the discriminator. One major advantage of SRGAN  is that the discriminator can help extract representative features from high-resolution data and enforce such features in the reconstructed images. Several variants of SRGAN are given in Refs.~\cite{chen2017fsrnet, wang2018recovering, wang2018esrgan, karras2018progressive, gan8759375, cheng2021mfagan, Long2021}.

The Transformer~\cite{vaswani2017attention} has revolutionized natural language processing (NLP) by introducing self-attention mechanisms, allowing it to efficiently process long-range dependencies in the sequences of data. This method can effectively capture contextual information from the entire input sequence, leading to significant advancements in various NLP tasks like machine translation, sentiment analysis and etc. The Transformer has also been introduced into the SISR problem~\cite{parmar2018image, fang2022cross,yang2020learning, lu2022transformer,  fang2022hybrid, wang2022detail, zou2022self, liang2022light}.  For example, Yang et al.~\cite{yang2020learning} developed the TTSR model, which uses a learnable texture extractor to extract textures from low-resolution (LR) images and reference high-resolution (HR) images in order to recover target HR images. Lu et al.~\cite{lu2022transformer} developed the ESRT model, which optimizes the original Transformer to achieve competitive reconstruction performance with low computational cost.

\subsection{Super-Resolution for Turbulent Flows} 

There is a significant  interest in developing SR techniques for high-resolution flow reconstructions. Fukami et al.~\cite{fukami2019super, Fukami_2020,liu2020deep} created an improved CNN-based hybrid DSC/MS model to explore multiple scales of turbulence and capture the spatio-temporal turbulence dynamics.  Liu et al.~\cite{liu2020deep}  developed another CNN-based model MTPC to simultaneously include spatial and temporal information to fully capture features in different time ranges. Xie et al. \cite{xie2018tempogan} introduced tempoGAN, which augments a GAN model with an additional discriminator network along with new loss functions that preserve temporal coherence in the generation of physics-based simulations of fluid flow.  Deng et al.~\cite{Deng2019SuperresolutionRO} demonstrated that both SRGAN and ESRGAN~\cite{wang2018esrgan} can produce good reconstructions. Yang et al.~\cite{yang2023super} created an FSR model based on a back-projection network to achieve 3D reconstruction. Xu et al.~\cite{xu2023super} introduced a Transformer-based SR method to build the SRTT model for capturing small-scale details of turbulent flow.

\subsection{Physics-Guided Machine Learning} 

Recent studies have shown promise in integrating physics into machine learning models for improved predictive performance~\cite{willard2022integrating}.  These methods typically enforce physics in the loss function~\cite{he2023physics,hanson2020predicting,karpatne2017physics,jia2019sdm2,read2019process,chen2021reconstructing} or use simulated data for pre-training and augmentation~\cite{jia2023physics,chen2023physics,he2023physics,chen2022physics,liu2022kgml}.  Hanson et al.~\cite{hanson2020predicting} introduced ecological principles as physical constraints into the loss function to improve the lake surface water phosphorus prediction. Karpatne et al.~\cite{karpatne2017physics} developed a hybrid machine learning and physics model to guarantee that the density of water at a lower depth is always greater than the density at any depth above.  Jia et al.~\cite{jia2019sdm2} and Read et al.~\cite{read2019process}  extended this idea by including an additional penalty for the violation of the energy conservation law. In the flow data reconstruction, Chen et al.~\cite{chen2021reconstructing} constructed a PGSRN method to enforce zero divergences of the velocity field in incompressible flows.  Despite the promise of these methods, they may lead to slow convergence in optimization and performance degradation, especially when the physical relationships are complex or have uncertain parameters. 

A means of imposing the physics is by considering the partial differential equations (PDEs) that govern the physical phenomena. In some cases, however, direct integration of the governing PDEs using standard numerical methods ~\cite{enwiki:1126400243} can become prohibitively expensive. An alternative is to solve PDEs via neural operators~\cite{li2020fourier, equer2023multi, boussif2022magnet}. For example, Li et al.~\cite{li2020fourier} introduced the Fourier neural operator (FNO) to model PDEs for learning the mappings between infinite-dimensional spaces of functions using the integral operator. The integral operator of this approach is restricted to convolution and instantiated through a linear transformation in the Fourier domain. However, the major limitation of neural operators for flow data reconstruction lies in their lack of explicit knowledge about the specific form of the underlying PDE (Naiver-Stoke equation). Neural operators directly learn the relation between input data and outputs without incorporating the intrinsic structure and physics encoded in the PDE. This can lead to inefficiencies and challenges in effectively capturing complex flow dynamics. An alternative direction is to embed the physics equations or relationships in the modeling structure~\cite{khandelwal2020physics, muralidhar2020phynet, bao2022physics}. One such example is the encoding of the Navier-Stokes equation in a recurrent unit, as demonstrated in our previous work~\cite{bao2022physics}. However, this method may accumulate errors in long-term predictions, and it does not consider the use of LES data in reconstructing DNS data within the recurrent unit.  




\section{Problem Under Investigation}

In this work, the transport of unsteady, three-dimensional turbulent flows is the subject of main consideration.  In all cases, the flow is assumed to be Newtonian and incompressible with a constant density.  In the formulation, the space coordinate is identified by the vector ${\bf x}\equiv {x,\ y,\ z}$,  and the time is denoted by $t$.  
The velocity field is denoted by ${\bf V} ({\bf x},t)$, with its three components ${u ({\bf x},t) ,\ v({\bf x},t),\ w({\bf x},t)}$ along the three flow directions ${\ x,\ y,\ z}$, respectively.  The pressure, the density, and the dynamic viscosity are denoted by  $p ({\bf x},t)$,  $\rho ({\bf x},t)$, and  $\nu$, respectively.  The latter two are assumed constant.  The (dummy) parameters $\textbf{Q} ({\bf x},t)$ (as a vector), and/or $Q$ (as a scalar) are used to denote a transport variable.

All of the flows considered are statistically homogeneous.  High-resolution DNS and lower-resolution LES data are considered on  $N_x \times N_y \times N_z$,  and   $M_x \times M_y \times M_z$ grid points, respectively. A box filter ~\cite{jing2000fuzzy} is employed to create the  LES data from the original DNS.  All of the statistical averages, including the Reynolds-averaged values are obtained by data ensembled over the entire domain. In this way, the ensemble averages, denoted by an over-bar are defined by:
\begin{equation}
    \overline{\bf{Q}(t)}=\frac{1}{N_x \times N_y \times N_z} \sum_i^{N_x}\sum_j^{N_y}\sum_k^{N_z} {\bf Q}(i,j,k,t),
\end{equation}
suitable for homogeneous flows.  In the training process, the available DNS data is at a regular time interval $\delta$, as $\textbf{Q}^d =\{\textbf{Q}^d(t)\}$ within the time $\{t_0,t_0+\delta,\ \dots, t_0+K\delta\}$. The objective is to predict high-resolution DNS data after the historical data, at time $\{t_0+(K+1)\delta, $\dots$, t_0+M\delta\}$. The variable $\textbf{Q}^l(x,y,z,t)$  represents the low-resolution LES data at time step $t$. Since the LES data can be created at a lower computational cost, they are used for both training and testing periods and at a higher frequency. The variable $\textbf{Q}^l = \{\textbf{Q}^l(t)\}$ denotes LES data within the time range $[t_0,t_0+M\delta]$. 

\begin{figure} [!h]
\centering
\includegraphics[width=0.7\linewidth]{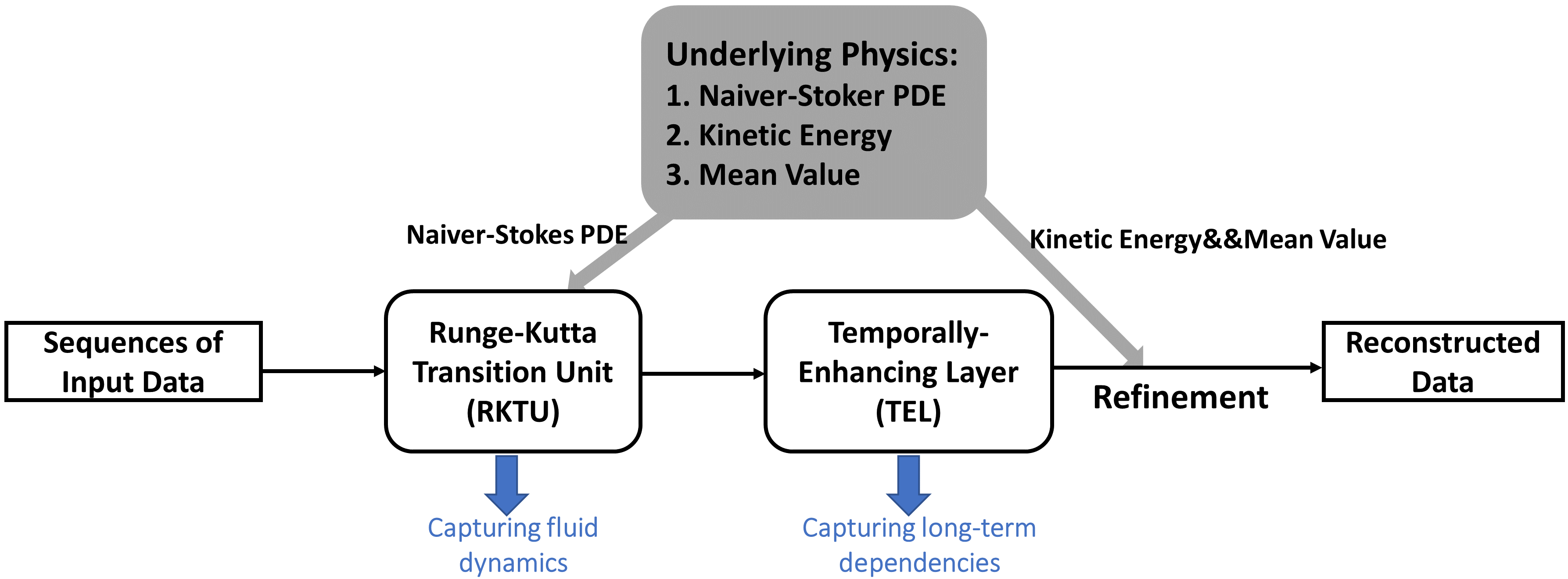}
\caption{The overall structure of the CNDE method.} 
\label{fig:overall_struc}
\end{figure}

The ``continuous networks using differential equations''  (CNDE) framework consists of two structural components: the Runge-Kutta transition unit (RKTU), and the temporally-enhance layer (TEL).  The training is done in two phases:  supervised super-resolution training, and degradation-based refinement.  These are shown in Fig.~\ref{fig:overall_struc}, and are described in order below.

\subsection{Runge-Kutta Transition Unit (RKTU)}

The data sets $\textbf{Q}$ pertaining to turbulent flows 
consist of the transport variables that interact with each other and evolve temporally and spatially.  The traditional temporal models, e.g., long-short term memory (LSTM)~\cite{LSTM}, rely on large and consecutive training samples to capture the underlying patterns over time. However, the amount of high-fidelity DNS data is often limited. The RKTU structure is developed for reconstructing flow variables over a long period, given an initial DNS sample $\textbf{Q}^d$ at $t$, and frequent low-resolution LES data samples  $\textbf{Q}^l$. The prediction follows an auto-regressive process in which the predicted DNS $\hat{\textbf{Q}}^d(x,y,z,{t})$ at time $t$, and frequent LES data $\textbf{Q}^l$ from the current time to the next interval [$t$,$t+\delta$] are used to predict the DNS at next time step $\hat{\textbf{Q}}^d(x,y,z,{t+\delta})$.

\begin{figure} [!h]
\centering
\includegraphics[width=0.6\linewidth]{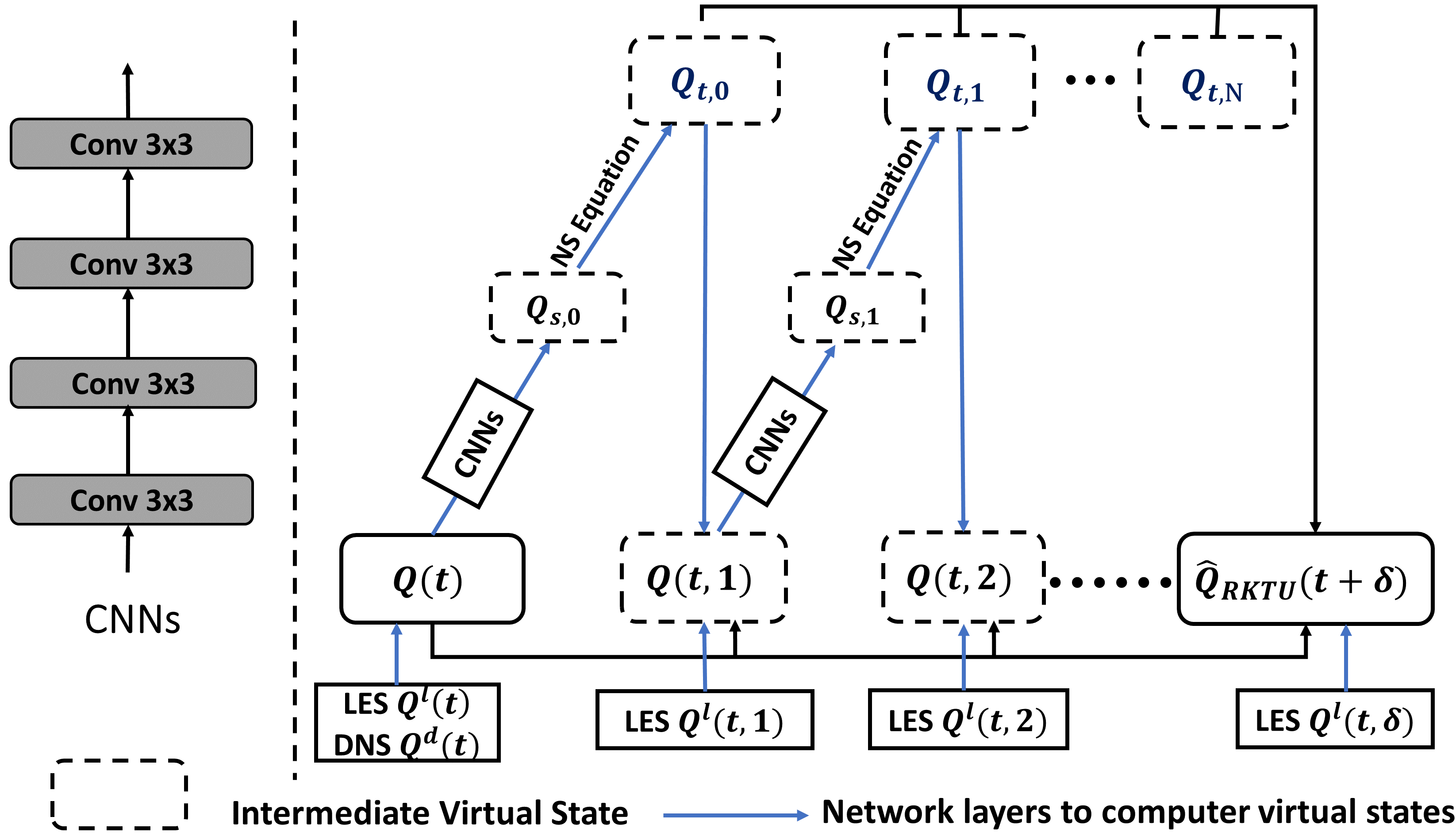}
\caption{The RKTU based on Naiver Stoke equation for reconstructing turbulent flow data in the spatio-temporal field. $\textbf{Q}_{s,n}$ and $\textbf{Q}_{t,n}$ denote the spatial and temporal derivatives, respectively, at each intermediate time step.} 
\label{fig:RKTU}
\end{figure}

The RKTU is based on the  Runge–Kutta (RK) discretization method~\cite{butcher2007runge}. The principal idea is to leverage the continuous physical relationship described by the underlying PDE to bridge the gap between the discrete data samples and the continuous flow dynamics. The scheme can be applied to any dynamical systems governed by deterministic PDEs. Consider the PDE of the target variables $\textbf{Q}$ as expressed by:  
\begin{equation}
\textbf{Q}_t = {\textbf{f}}(t, \textbf{Q};\theta), 
\end{equation}
where $\textbf{Q}_t$ denotes  the temporal derivative of $\textbf{Q}$,  and ${\textbf{f}}(t, \textbf{Q};\theta)$ is a non-linear function (parameterized by coefficient $\theta$) that summarizes the current value of $\textbf{Q}$ and its spatial variations. The turbulence data follows the Navier-Stokes equation for an incompressible flow.  Thus for  
${\bf Q} \equiv {\bf V}({\bf x},t)$:
$$\nabla \cdot {\bf Q}=0,$$
\begin{equation}
{\textbf{f}(\textbf{Q})} = -\frac{1}{\rho} \nabla p + \nu \Delta \textbf{Q} - (\textbf{Q}\cdot \nabla) \textbf{Q},
\label{eq:NS}
\end{equation} 
The term $\nabla$ denotes the gradient operator and $\Delta =\nabla \cdot \nabla $ on each of the components of the velocity. The independent variable $t$ is omitted in the function ${\textbf{f}}(\cdot)$ because ${\textbf{f}}(\textbf{Q})$ in the Navier-Stokes equation is for a specific time $t$ (same with $t$ in $\textbf{Q}_t$). Figure \ref{fig:RKTU} shows the overall structure of the method and involves a series of intermediate states $\{\textbf{Q}(t,0),\textbf{Q}(t,1),\textbf{Q}(t,2),\dots,\textbf{Q}(t,N)\}$.  The  temporal gradients are estimated at these states $\{\textbf{Q}_{t,0},\textbf{Q}_{t,1},\textbf{Q}_{t,2},\dots,\textbf{Q}_{t,N}\}$. 
Starting from $\textbf{Q}(t,0)=\textbf{Q}(t)$, the RKTU 
estimates the temporal gradient as  $\textbf{Q}_{t,0}$, and then moves  $\textbf{Q}(t)$ towards the gradient direction to create the next intermediate state $\textbf{Q}(t,1)$. The process is repeated for  $N$  intermediate states. For the fourth-order RK method, as employed here, $N=3$. 

For the starting data point $\textbf{Q}(t)$,  an augmentation mechanism is adopted by combining the DNS and LES data: $\textbf{Q}(t) = W^d \textbf{Q}^d (t) + W^l\textbf{Q}^l (t)$, where $W^d$ and $W^l$ are trainable model parameters, and $\textbf{Q}^l(t)$ is the up-sampled LES data with the same resolution as DNS. The RKTU estimates the first temporal gradient $\textbf{Q}_{t,0}=\textbf{f}(\textbf{Q}(t))$ using the Navier-Stokes equation and computes the next intermediate state variable $\textbf{Q}(t,1)$ by moving the flow data $\textbf{Q}(t)$ along the direction of temporal derivatives. Given frequent LES data, the intermediate states $\textbf{Q}(t,n)$ are also augmented by using LES data $\textbf{Q}^l(t,n)$, as $\textbf{Q}(t,n) = W^d \textbf{Q}(t,n) + W^l \textbf{Q}^l(t,n)$, and they follow the same process to move $\textbf{Q}(t)$ along the estimated gradient $\textbf{Q}_{t,n}$ to compute the next intermediate states $\textbf{Q}(t,n+1)$.  

\begin{equation}    \label{eq:RK-steps}
\begin{aligned}
\textbf{Q}(t,{1}) &= \textbf{Q}(t) +  \delta \frac{\textbf{Q}_{t,0}}{2},\\
\textbf{Q}(t,{2}) &= \textbf{Q}(t) + \delta\frac{\textbf{Q}_{t,1}}{2},\\
\textbf{Q}(t,{3}) &= \textbf{Q}(t) + \delta\textbf{Q}_{t,2}.
\end{aligned}
\end{equation}

The temporal derivative $\textbf{Q}_{t,3}$ is then computed from the last intermediate state by $\textbf{f}(\textbf{Q}(t,{3}))$.  
According to Eq.~(\ref{eq:RK-steps}), the intermediate LES data $\textbf{Q}^l(t,n)$ are selected as $\textbf{Q}^l(t,1)=\textbf{Q}^l(t+\delta/2)$, $\textbf{Q}^l(t,2)=\textbf{Q}^l(t+\delta/2)$, and $\textbf{Q}^l(t,3)=\textbf{Q}^l(t+\delta)$. Finally, RKTU combines all the intermediate temporal derivatives as a composite gradient to calculate the final prediction of  next step flow data $\hat{\textbf{Q}}_\text{RKTU}(t+\delta)$: 
\begin{equation}    \label{eq:RK-training_comb}
\hat{\textbf{Q}}_\text{RKTU}(t+\delta)= \textbf{Q}(t) + \sum_{n=0}^N w_n \textbf{Q}_{t,n}, 
\end{equation}
where $\{w_n\}_{n=1}^N$ are the trainable model parameters.  

The RKTU requires the temporal derivatives in the Navier-Stokes equation. 
The RKTU estimates the temporal derivatives through the function ${\textbf{f}}(\cdot)$. According to  Eq.~(\ref{eq:NS}), the evaluation of ${\textbf{f}}(\cdot)$ requires explicitly  
estimation of the first-order and second-order spatial derivatives. 
One of the most popular approaches for evaluating spatial derivatives is through finite difference methods (FDMs)~\cite{enwiki:1126400243}. However, the discretization in FDMs can cause larger errors for locations with complex dynamics. 
The RKTU structure, as depicted in Fig.~(\ref{fig:RKTU}), utilizes convolutional neural network layers (CNNs) to replace the FDMs. The CNNs have the inherent capability to learn additional non-linear relationships from data and capture the spatial derivatives required in the Navier-Stokes equation. After estimating the first-order and second-order spatial derivatives, they are used in  Eq.~(\ref{eq:NS}) to obtain the temporal derivative $\textbf{Q}_{t,n}$.

The padding strategies for CNNs also need to be considered. Standard padding strategies (e.g., zero padding) do not satisfy the spatial boundary conditions of the flows considered here.  These conditions describe how the flow data interact with the external environment. With the assumption of homogeneous turbulence, periodic boundary conditions are imposed on all three flow directions. Thus, periodic data augmentation is made for each of the $6$ faces (of the 3D cubic data) with an additional two layers of data before feeding it to the model. 




\begin{figure*} [!h]
\centering
\subfigure[Enhancing Method]{ \label{fig:a}{}
\includegraphics[width=0.6\linewidth]{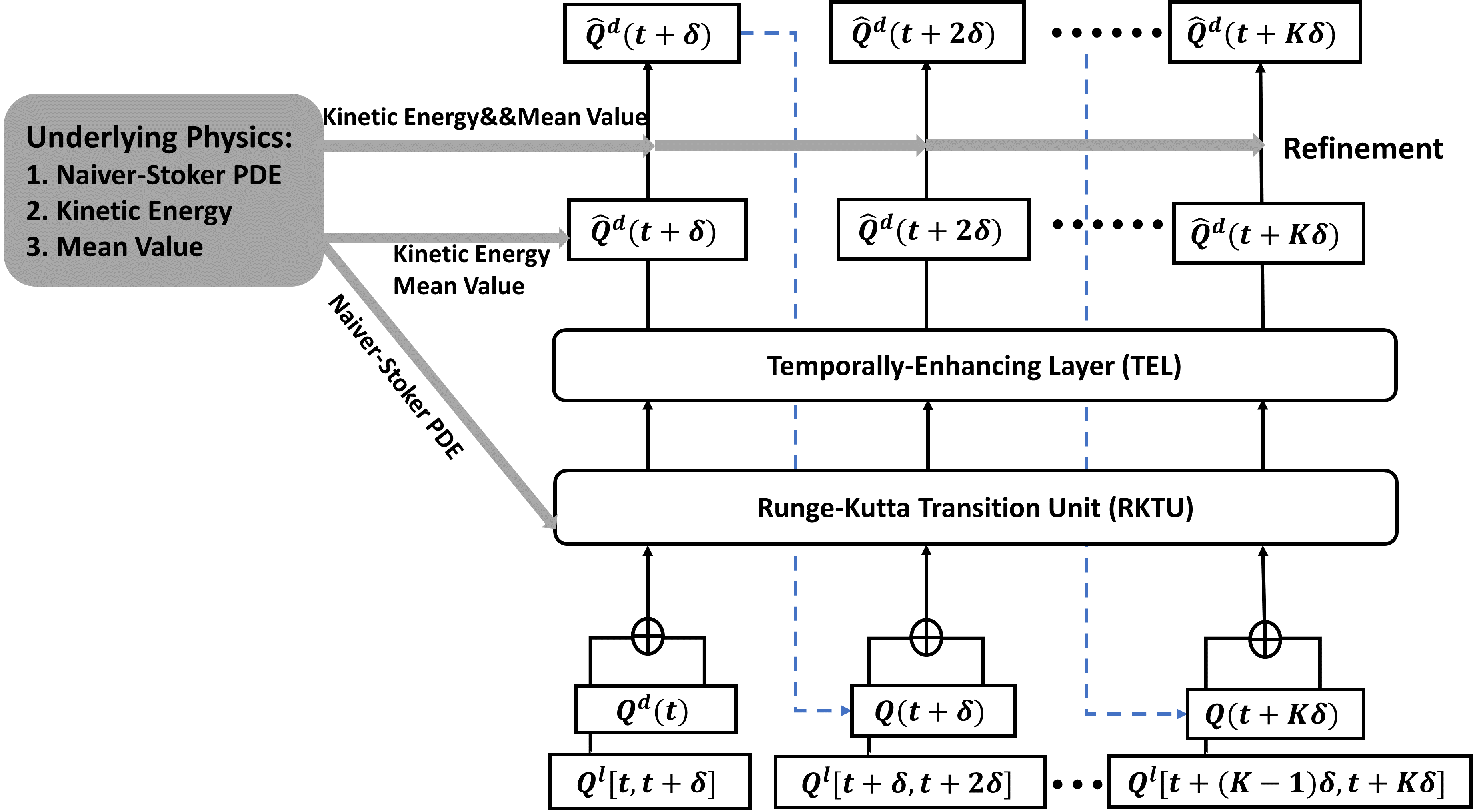}
}
\\
\subfigure[Residual Learning Method]{ \label{fig:b}{}
\includegraphics[width=0.6\linewidth]{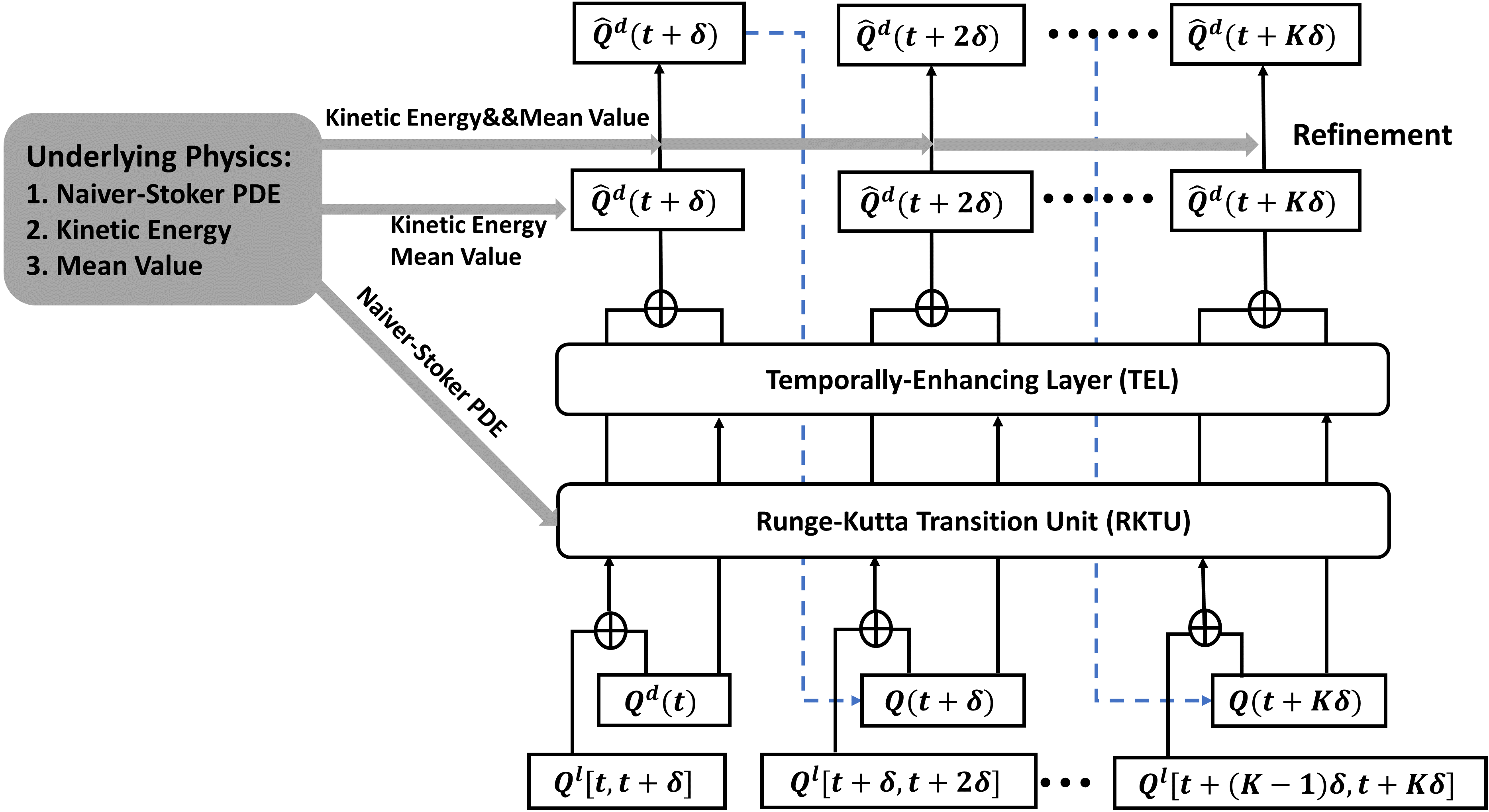}
}
\caption{Details of ``continuous networks using differential equations'' (CNDE) for reconstructing  $\textbf{Q}$. Parts (a) and (b) depict two different approaches to incorporating the TEL component. The solid lines represent the forwarding process in CNDE-based methods. The blue dashed lines represent the reconstructed flow data $\hat{\textbf{Q}}^d$ as used as input for the reconstruction of the next time step.} 
\label{fig:Framework}
\end{figure*}

\subsection{Temporally-Enhancing Layer (TEL)}

The RKTU can capture the data in the spatial and temporal field between a pair of consecutive data points, but it may cause large reconstruction errors in the long-time prediction if the time interval $\delta$ is large. Temporal models,  such as long-short-term memory (LSTM)~\cite{LSTM}, and temporal convolutional network (TCN)~\cite{lea2016temporal} are widely used to capture the long-term dependencies in time series prediction. In this case, the LSTM  model is incorporated in a temporally-enhancing layer (TEL) to further enhance  the RKTU to capture long-term temporal dependencies. This TEL structure can be replaced by other existing temporal models such as TCN. Figure \ref{fig:Framework}  
shows two different approaches for integrating the TEL structure with the RKTU structure. 
In the first enhancing method shown in Fig.~\ref{fig:Framework} (a), the RKTU output flow data $\hat{\textbf{Q}}_\text{RKTU}$ 
is fed to the TEL structure, which is essentially an LSTM layer. After further processing through the TEL structure, the model produces the reconstructed flow data 
$\hat{\textbf{Q}}^\text{d}(t)$. Given the true DNS data  $\textbf{Q}^d (t)$ in the training set, the reconstructed loss $\mathcal{L}_\text{recon}$ can be expressed using the mean squared error (MSE) loss:
\begin{equation}
    \mathcal{L}_{\text{recon}} = \sum_t\text{MSE}(\hat{\textbf{Q}}^d(t),\textbf{Q}^d(t)).
    \label{eq:enhancing}
\end{equation}

The second method  uses the TEL structure to complement the output of the RKTU structure, i.e., learning the residual of the RKTU  output, as shown in Fig.~\ref{fig:Framework} (b). 
In the training process, both true DNS data $\textbf{Q}^d$ at time $\{t,\dots t+(K-1)\delta\}$ and RKTU output $\hat{\textbf{Q}}_\text{RKTU}$ are used to produce the corresponding temporal output feature $\hat{\textbf{Q}}_{\text{TEL}}$ at time  $\{t+\delta, \dots, t+K\delta\}$. Then in the testing process,  this method uses only the initial true DNS data $\textbf{Q}^d$ in time $t+K\delta$ and the next series of predicted DNS data $\hat{\textbf{Q}}^d$ as the DNS input to generate $\hat{\textbf{Q}}_{\text{TEL}}$.  Finally, 
this method adopts a linear combination to combine the RKTU output $\hat{\textbf{Q}}_\text{RKTU}$ and corresponding TEL output $\hat{\textbf{Q}}_{\text{TEL}}$ to obtain the final reconstructed output $\hat{\textbf{Q}}^d$, which can be represented as:
\begin{equation}
    \hat{\textbf{Q}}^d(t) = w_{r}^t \hat{\textbf{Q}}_\text{RKTU}(t) + w_{t}^t \hat{\textbf{Q}}_\text{TEL}(t) 
    \label{eq:attention},
\end{equation}
where $w_{r}^t$ and $w_{t}^t$ are trainable parameters. Finally, the reconstructed loss $\mathcal{L}_\text{recon}$ can also be represented by Eq.~(\ref{eq:enhancing}).

\subsection{Physical Constraints and Refinements}

\subsubsection{Physical Constraints}

For a more accurately reconstructed field, some additional constraints are imposed on the data.  Two such constraints are imposed  by considering the consistency of (i) the mean velocity field, and (ii) the kinetic energy of turbulence.   For (i), the loss function $\mathcal{L}_\text{mean}$ between reconstructed data $\hat{\textbf{Q}}^d$ and true DNS data $\textbf{Q}^d$ is considered:
\begin{equation}
    \mathcal{L}_{\text{mean}} = |\overline{\textbf{Q}^d} -\overline{\hat{\textbf{Q}}^d}|.
    \label{eq:mean}
\end{equation}
For (ii),  the kinetic energy,],  defined as:   
\begin{equation}
    \mathcal{K} = \frac{1}{2}(u^2 + v^2 + w^2),
\end{equation}
is monitored.  For this, the loss function  is $\mathcal{L}_{\text{kinetic}}$ is:

\begin{equation}
    \mathcal{L}_{\text{kinetic}} = |\mathcal{K}(\textbf{Q}^d)-\mathcal{K}(\hat{\textbf{Q}}^d)|,
    \label{eq:kinetic}
\end{equation}
where $\mathcal{K}(\textbf{Q}^d)$ and $\mathcal{K}(\hat{\textbf{Q}}^d)$ denote the kinetic energy  of $\textbf{Q}^d$ and $\hat{\textbf{Q}}^d$, respectively. The  overall  loss function $\mathcal{L}$ is: 

\begin{equation}
    \mathcal{L} = \alpha_{0} \mathcal{L}_{\text{recon}} + \alpha_{1} \mathcal{L}_{\text{mean}}+
    \alpha_{2} \mathcal{L}_{\text{kinetic}}, 
\end{equation}
is considered in which $\alpha_{0}$, $\alpha_{1}$, and $\alpha_{2}$ represent the hyperparameters to control the balance amongst the three constituents.

\subsubsection{Degradation-Based Refinement}

As shown in Fig.~\ref{fig:Framework}, the scheme preserves the physical constraints in the training process and also employs these constraints in the degradation-based test-time refinement process. The objective is to mitigate accumulated errors and structural distortions over long-term prediction by enforcing the physical consistency. The  
 refinement process includes the same set of the loss function: the degradation loss $\mathcal{L}_{\text{deg}}$, the equal-mean loss $ \mathcal{L}^{'}_{\text{mean}}$,  and the kinetic energy loss $ \mathcal{L}^{'}_{\text{kinetic}}$ loss. 
Since it is not possible to access true DNS data during the testing phase, the difference between true DNS $\textbf{Q}^d$ and the reconstructed data $\hat{\textbf{Q}}^d$ cannot be directly minimized. Thus, in order to protect the overall structure of flow data, a reverse degradation process is employed by using a separate convolutional network, for mapping reconstructed data $\hat{\textbf{Q}}^d$ to the corresponding low-resolution LES data $\hat{\textbf{Q}}^l$. The loss $\mathcal{L}_{\text{deg}}$ between $\hat{\textbf{Q}}^l$ and real LES data $\textbf{Q}^l$ is:
\begin{equation}
    \mathcal{L}_{\text{deg}} = \text{MSE}(\hat{\textbf{Q}}^l,\textbf{Q}^l).
    \label{eq:enhancing_test}
\end{equation}

Also, the mean values from the true DNS cannot be used in the equal-mean loss function. Therefore,  the corresponding  values from the LES data are used as an approximation.   As such, the equal-mean loss $ \mathcal{L}^{'}_{\text{mean}}$ between the reconstructed flow data $\hat{\textbf{Q}}^d$ and the true LES data $\textbf{Q}^l$ can be directly minimized:
\begin{equation}
    \mathcal{L}^{'}_{\text{mean}} = |\overline{\textbf{Q}^l} -\overline{\hat{\textbf{Q}}^d}|,
    \label{eq:mean_test}
\end{equation}

Similarly, the exact kinetic energy of flow data is not available during the testing period.  These values are  taken from the DNS in the training data:  
\begin{equation}
    \mathcal{L^{'}}_{\text{kinetic}} = |\mathcal{K}(\hat{\textbf{Q}}^d)-\tilde{\mathcal{K}}|.
    \label{eq:kinetic_test}
\end{equation}

The final refinement loss function
is in the same format $\mathcal{L}{'} = \alpha_{0} \mathcal{L}_{\text{deg}} + \alpha_{1} \mathcal{L{'}}_{\text{mean}}+
    \alpha_{2} \mathcal{L}{'}_{\text{kinetic}}$.  
The loss $ \mathcal{L}^{'}$  is adopted to directly adjust the state of reconstructed data for 10 epochs at each test-time time step, and yield an improved reconstruction performance. 

\section{MODEL APPRAISAL}

\subsection{Flows Considered}
\label{sec:dataset}
To assess the performance of the proposed methodology, the data sets pertaining to two turbulent flows  are considered: a forced isotropic turbulent flow (FIT)~\cite{FITdata}, and
the Taylor-Green vortex (TGV)  ~\cite{brachet1984taylor} flow. In both cases, the mean velocity is zero, $\overline{\bf V}(t)=0$, and the  Reynolds number is large enough for the flow to exhibit turbulent characteristics.

The FIT data ~\cite{FITdata} is publicly available from the Johns Hopkins University. This dataset contains the original DNS of forced isotropic turbulence on a $1024 \times 1024 \times 1024 $ collocation points.   The flow is forced by injecting energy into the flow at small waver numbers.   The DNS data contains $5024$ time steps with time intervals of $0.002$s and includes both the velocity and the pressure fields. The original DNS data are downsampled to  $64 \times 64 \times 64$ grids.   The LES data are created on $16 \times 16 \times 16$ grids.  The loss $\mathcal{L}{'}_{\text{kinetic}}$ is not considered for this flow.

The Taylor-Green vortex (TGV)~\cite{brachet1984taylor}  is an incompressible flow.  The evolution of the TGV includes vorticity stretching and the consequent production of small-scale, dissipating eddies. A box flow, with a cubic periodic domain of $[-\pi,\pi]$ (in all three directions) is considered, with the initial conditions:
\begin{equation}
\begin{aligned}
u (x,y,z,0) &= \sin(x) \cos(y) \cos(z),\\ 
v(x,y,z,0) &= - \cos(x)\sin(y)\cos(z),\\ 
w(x,y,z,0) &= 0.   
\end{aligned}
\end{equation}

The LES and DNS resolutions are $ 32 \times 32 \times 65$ and $ 128 \times 128 \times 65  $ respectively. Both LES and DNS data are produced along the $65$ equally-spaced grid points along the $z$ axis. 

\subsection{Comparative Assessments}
\begin{table}[!t]
\small
\newcommand{\tabincell}[2]{\begin{tabular}{@{}#1@{}}#2\end{tabular}}
\centering
\caption{Reconstruction performance (measured by SSIM, and dissipation difference) on $(u,v,w)$ channels by different methods in the FIT dataset. The performance is measured by the average results of the first 10 time steps.}
\begin{tabular}{|l|cccc|}
\hline
\textbf{Method} & SSIM & Dissipation Difference \\ \hline 
SRCNN&(0.859, 0.851, 0.851) &(0.301, 0.303, 0.303)&\\ 
RCAN& (0.861, 0.859, 0.859)&(0.299, 0.301, 0.300)&\\ 
HDRN&(0.861, 0.860, 0.862) &(0.298, 0.298, 0.297)\\
FSR&(0.861, 0.860, 0.861) &(0.299, 0.297, 0.296)\\
DCS/MS&(0.861, 0.862, 0.862) &(0.298, 0.295, 0.294)&\\
SRGAN& (0.862, 0.861, 0.863) &(0.296, 0.294, 0.294)&\\  
FNO&(0.874, 0.875, 0.874) &(0.265, 0.266, 0.273)\\
CTN& (0.881, 0.880, 0.881) &(0.253, 0.254, 0.254)&\\  
\hline
RKTU&(0.898, 0.899, 0.898)&(0.260, 0.261, 0.259)&\\ 
CNDEp-E&(0.909, 0.909, 0.907) &(0.244, 0.243, 0.245)&\\
CNDEp-R&(0.904, 0.905, 0.905) &(0.249, 0.248, 0.248)&\\
CNDE-E&{(0.927, 0.921, 0.922)} &{(0.193, 0.194, 0.197)}&\\
CNDE-R&{(0.921, 0.919, 0.920)} &{(0.196, 0.196, 0.200)}&\\
\hline
\end{tabular}
\label{fig:table1}
\end{table}

\begin{table}[!t]
\small
\newcommand{\tabincell}[2]{\begin{tabular}{@{}#1@{}}#2\end{tabular}}
\centering
\caption{Reconstruction performance (measured by SSIM, and dissipation difference) on $(u,v,w)$ channels by different methods in the TGV dataset. The performance is measured by the average results of the first 10 time steps.}
\begin{tabular}{|l|cccc|}
\hline
\textbf{Method} & SSIM & Dissipation Difference$\times$10 \\ \hline 
SRCNN&(0.602, 0.603, 0.626) &(0.083, 0.087, 0.079)&\\ 
RCAN& (0.627, 0.622, 0.631)&(0.073, 0.074, 0.071)&\\ 
HDRN&(0.638, 0.638, 0.641) &(0.072, 0.072, 0.068)\\
FSR&(0.646, 0.648, 0.649) &(0.070, 0.073, 0.066)\\
DSC/MS&(0.647, 0.649, 0.649) &(0.070, 0.071, 0.065)&\\
SRGAN& (0.661, 0.658, 0.666) &(0.068, 0.067,0.058)&\\
FNO&(0.645, 0.646, 0.648) &(0.072, 0.071, 0.072)\\
CTN& (0.623, 0.624, 0.627) &(0.093, 0.096, 0.087)&\\  
\hline
RKTU&(0.708, 0.708, 0.688)&(0.049, 0.046, 0.043)&\\ 
CNDEp-E&(0.724, 0.723, 0.708)&(0.046, 0.041, 0.039)&\\ 
CNDEp-R&(0.720, 0.719, 0.701) &(0.046, 0.045, 0.040)&\\
CNDE-E&{(0.938, 0.918, 0.876)} &{(0.031, 0.032, 0.026)}&\\
CNDE-R&{(0.917, 0.909, 0.877)} &{(0.033, 0.034, 0.028)}&\\
\hline
\end{tabular}
\label{fig:table2}
\end{table}

\subsubsection{CNDE Method and Baselines}
The performance of the  CNDE method is evaluated and compared with several existing methods for image SR and turbulent flow downscaling.
Specifically, the proposed CNDE-based methods, CNDE-E (enhancing-based TEL method) and CNDE-R (residual learning-based TEL method)$\footnote{The source code is at \url{https://drive.google.com/drive/folders/15PhF_q1HcJpXZIvxnR1mMbd8hbkxBbT_?usp=share_link}}$, were implemented.  Additionally, four popular SR methods, namely SRCNN~\cite{dong2014learning}, RCAN~\cite{zhang2018image}, HDRN~\cite{Duong2021}, SRGAN~\cite{ledig2017photo}, two popular dynamic fluid downscaling methods, DCS/MS~\cite{fukami2019super} and FSR~\cite{yang2023super}, and Fourier neural operator (FNO)\cite{li2020fourier}, are  used as baselines.  To better verify the effectiveness of each of the model's components,  four additional baselines are included: convolutional transition network (CTN), RKTU, CNDEp-E, and CNDEp-R.  The CTN is created by combining SRCNN and LSTM~\cite{LSTM}. CNDEp-E and CNDEp-R are similar to CNDE-E and CNDE-R, but they are created without using the degradation-based refinement process. 

By comparing the CTN with the RKTU, the objective is to demonstrate the advantages of the RKTU in spatio-temporal DNS reconstruction. By comparing the RKTU with the CNDEp-based methods, the goal is to show the effectiveness of introducing the TEL structure. The advantages of the refinement process are demonstrated by comparing the CNDEp-based and  CNDE-based methods.

\begin{figure*} [!h]
\centering
\subfigure[{$u$ Channel.}]{ \label{fig:a}{}
\includegraphics[width=0.33\linewidth]{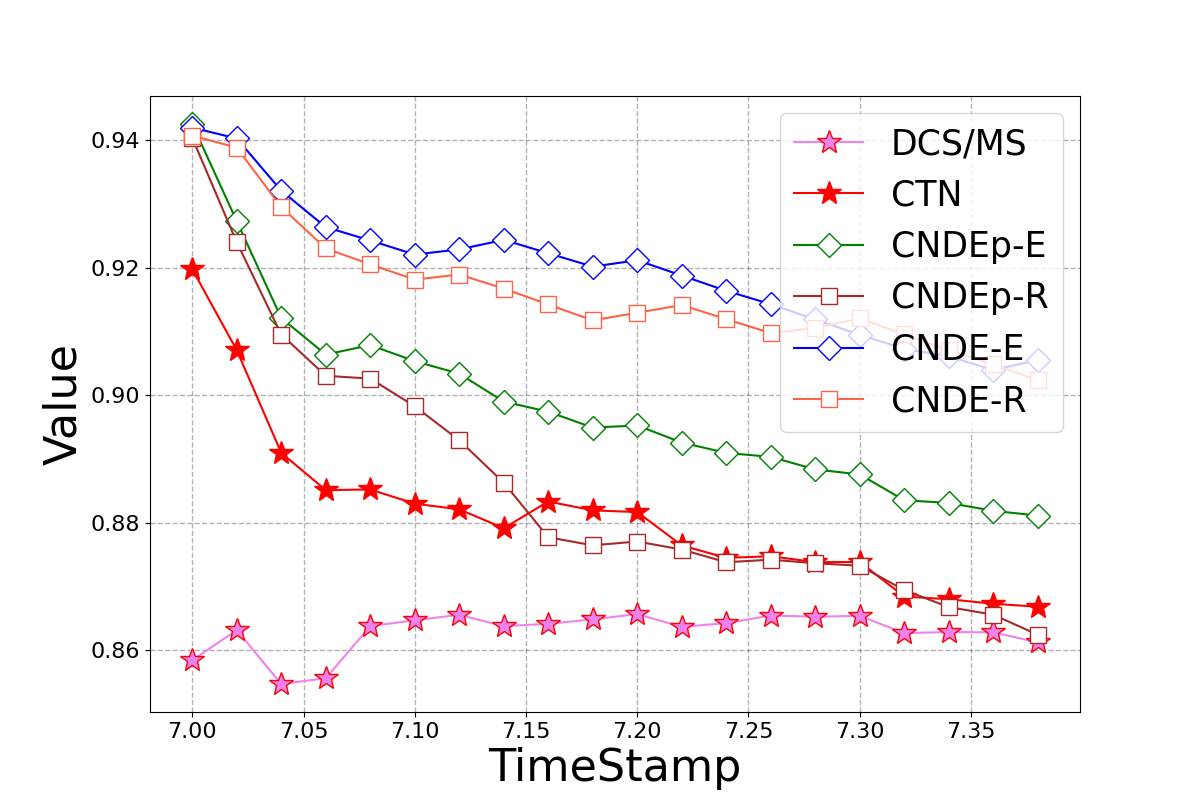}
}
\subfigure[{$v$ Channel.}]{ \label{fig:b}{}
\includegraphics[width=0.33\linewidth]{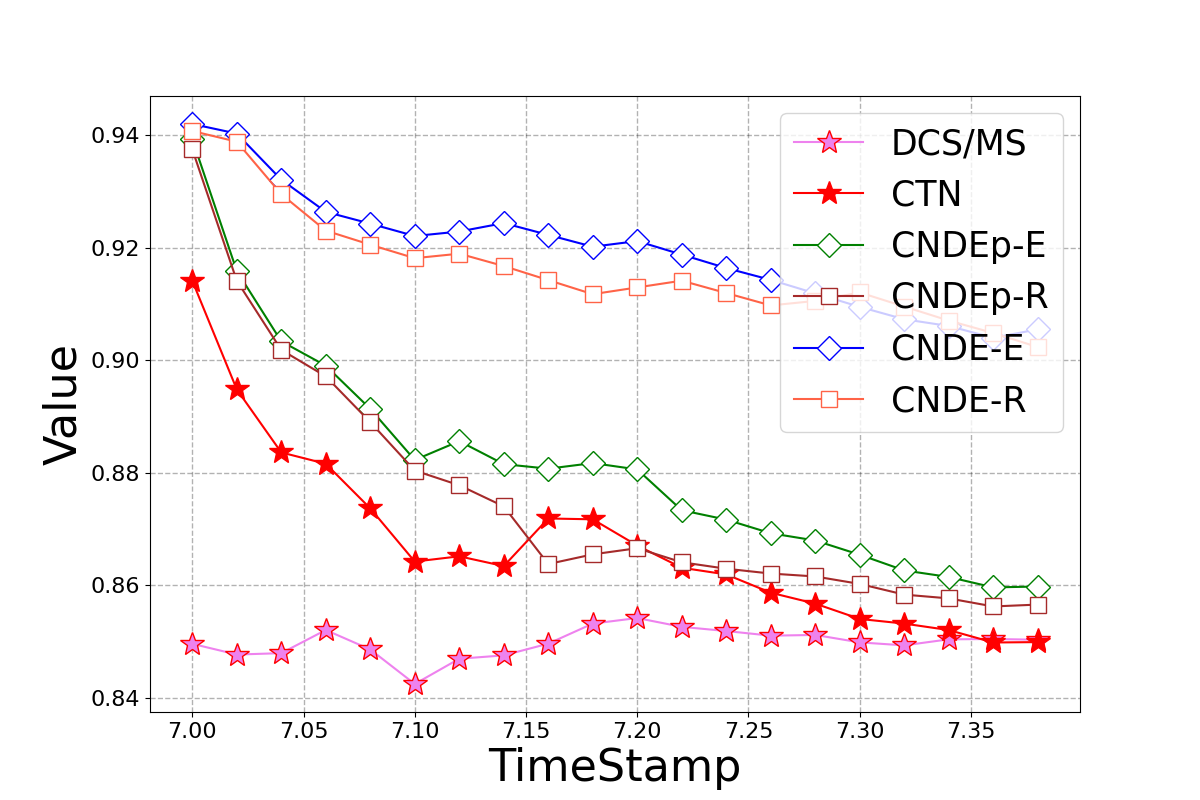}
}
\subfigure[{$w$ Channel.}]{ \label{fig:b}{}
\includegraphics[width=0.33\linewidth]{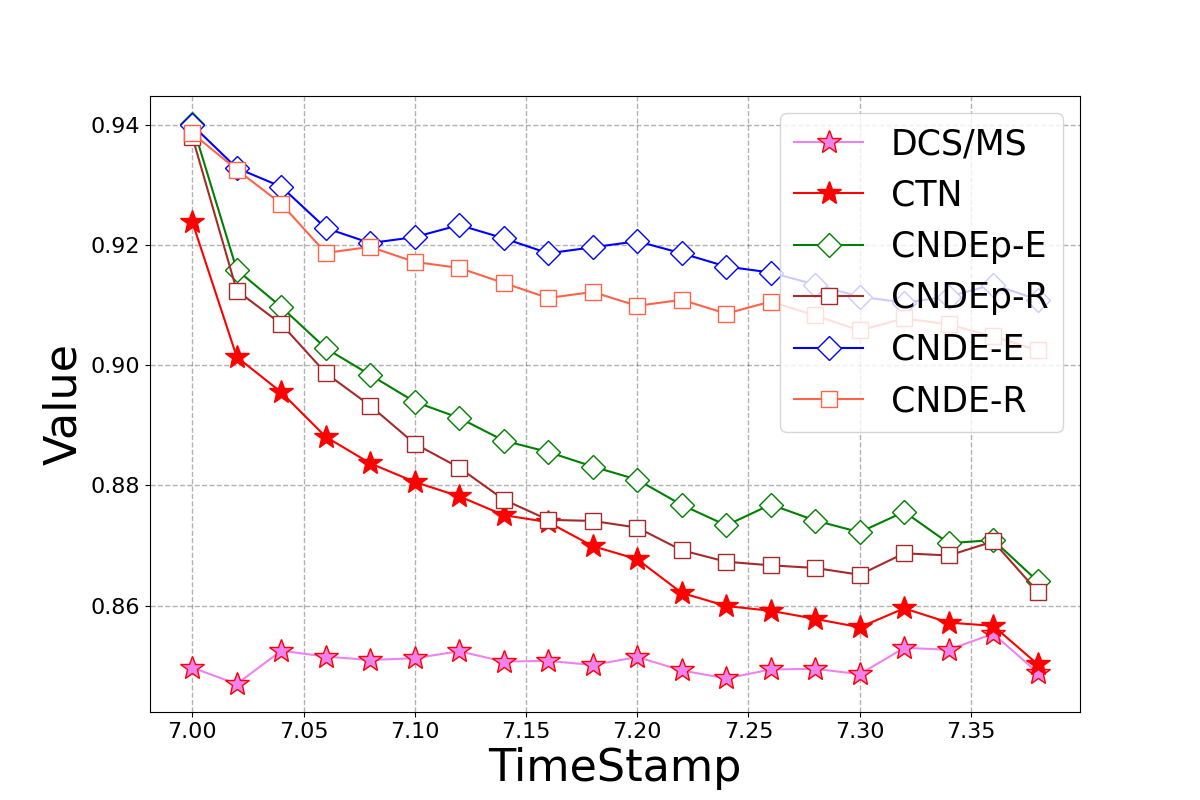}
}
\caption{Change of SSIM values produced by different models from 1st (7s) to 20th (7.4s) time step in the FIT dataset.}
\label{fig:tf_plot1}
\end{figure*}

\begin{figure*} [!h]
\centering
\subfigure[{$u$ Channel.}]{ \label{fig:a}{}
\includegraphics[width=0.33\linewidth]{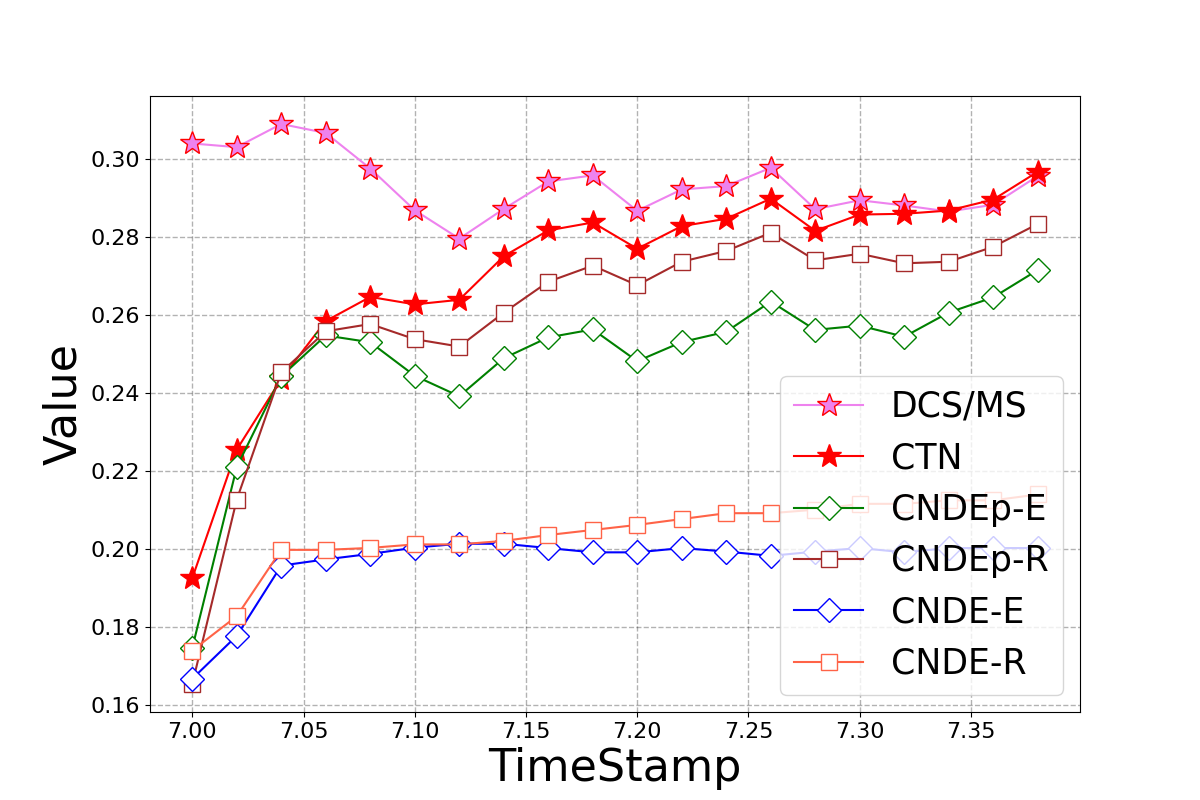}
}
\subfigure[{$v$ Channel.}]{ \label{fig:b}{}
\includegraphics[width=0.33\linewidth]{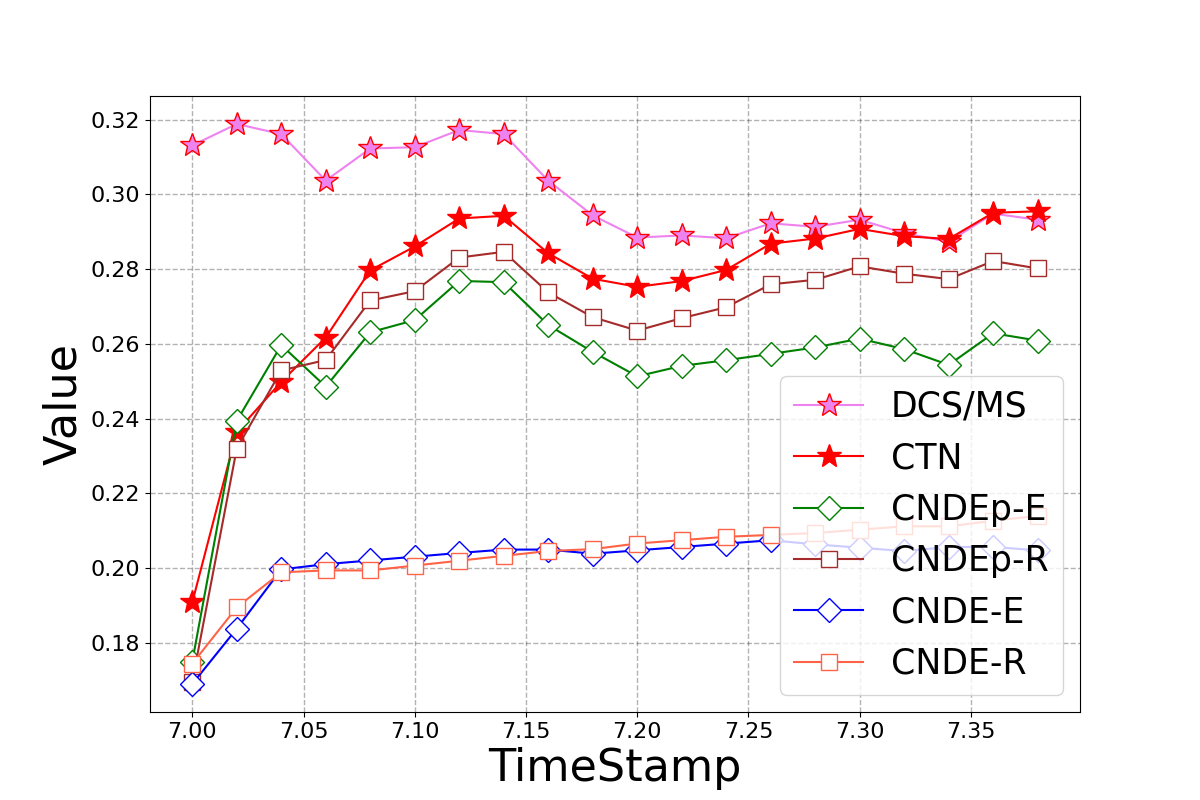}
}
\subfigure[{$w$ Channel.}]{ \label{fig:b}{}
\includegraphics[width=0.33\linewidth]{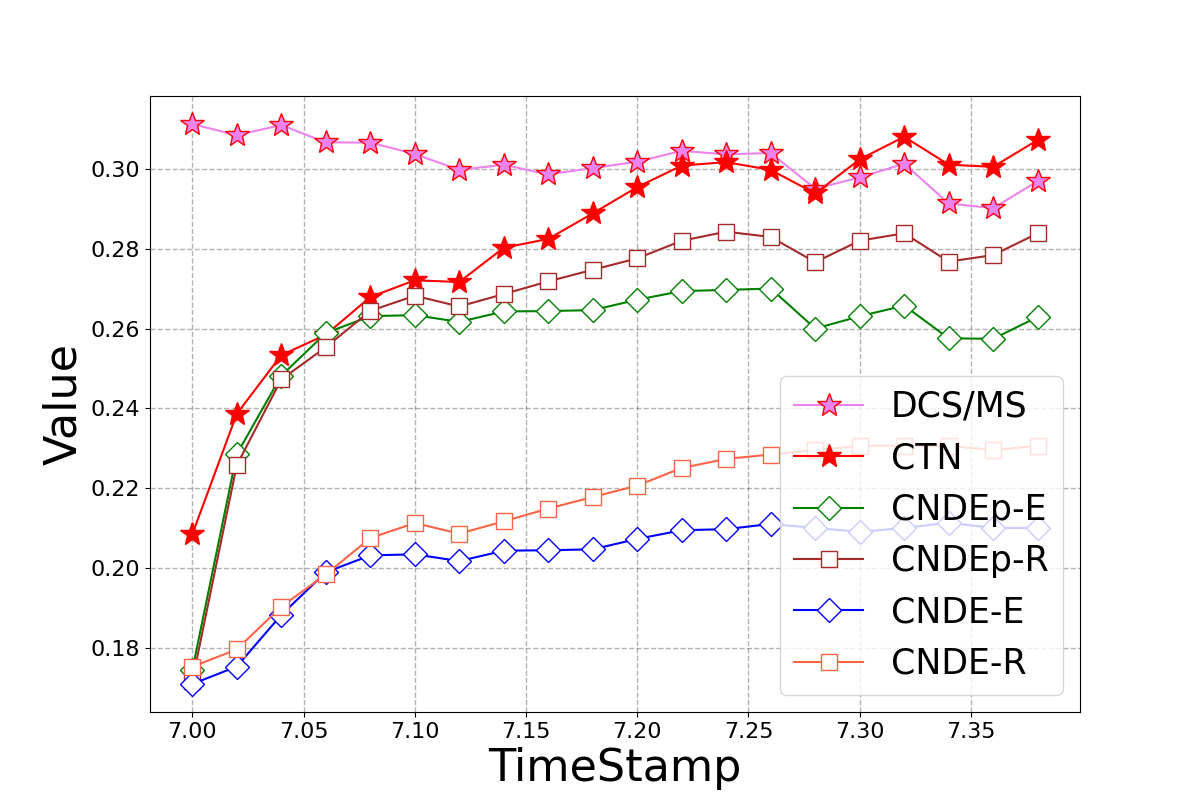}
}
\caption{Change of dissipation difference  by different models from 1st (7s) to 20th (7.4s) time step in the FIT dataset.}
\label{fig:tf_plot2}
\end{figure*}

\begin{figure*} [!h]
\centering
\subfigure[{$u$ Channel.}]{ \label{fig:a}{}
\includegraphics[width=0.32\linewidth]{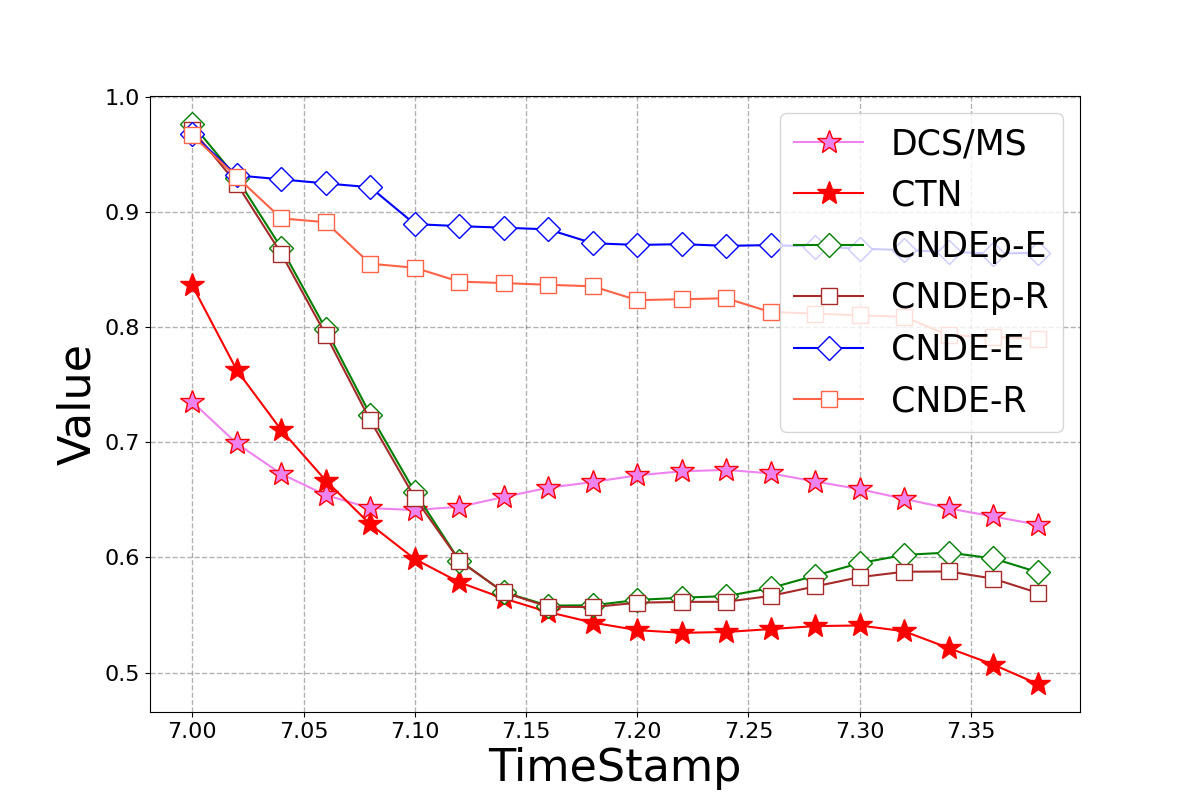}
}\hspace{-0.1in}
\subfigure[{$v$ Channel.}]{ \label{fig:b}{}
\includegraphics[width=0.32\linewidth]{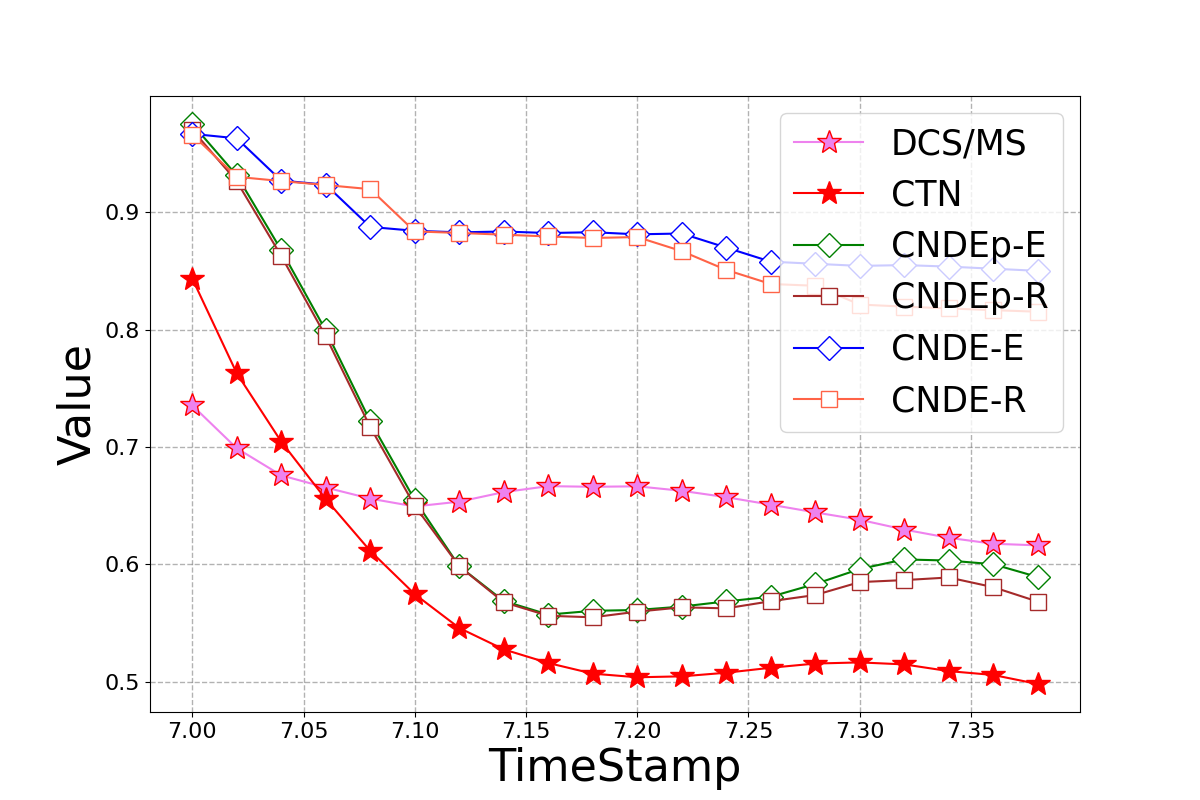}
}\hspace{-0.1in}
\subfigure[{$w$ Channel.}]{ \label{fig:b}{}
\includegraphics[width=0.32\linewidth]{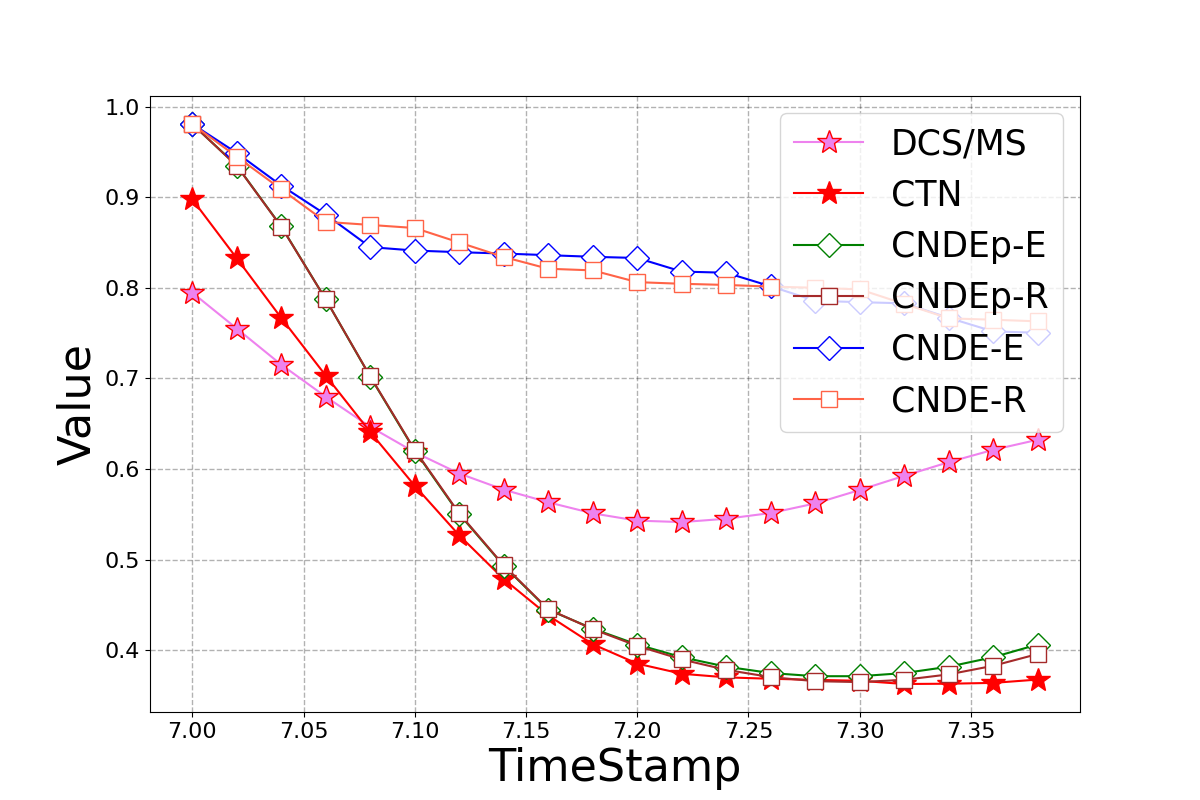}
}
\vspace{-.1in}
\caption{Change of SSIM values produced by different models from 1st (80s) to 20th (120s) time step in the TGV dataset.}
\label{fig:tf_plot4_tgv}
\end{figure*}

\begin{figure*} [!h]
\centering
\subfigure[{$u$ Channel.}]{ \label{fig:a}{}
\includegraphics[width=0.32\linewidth]{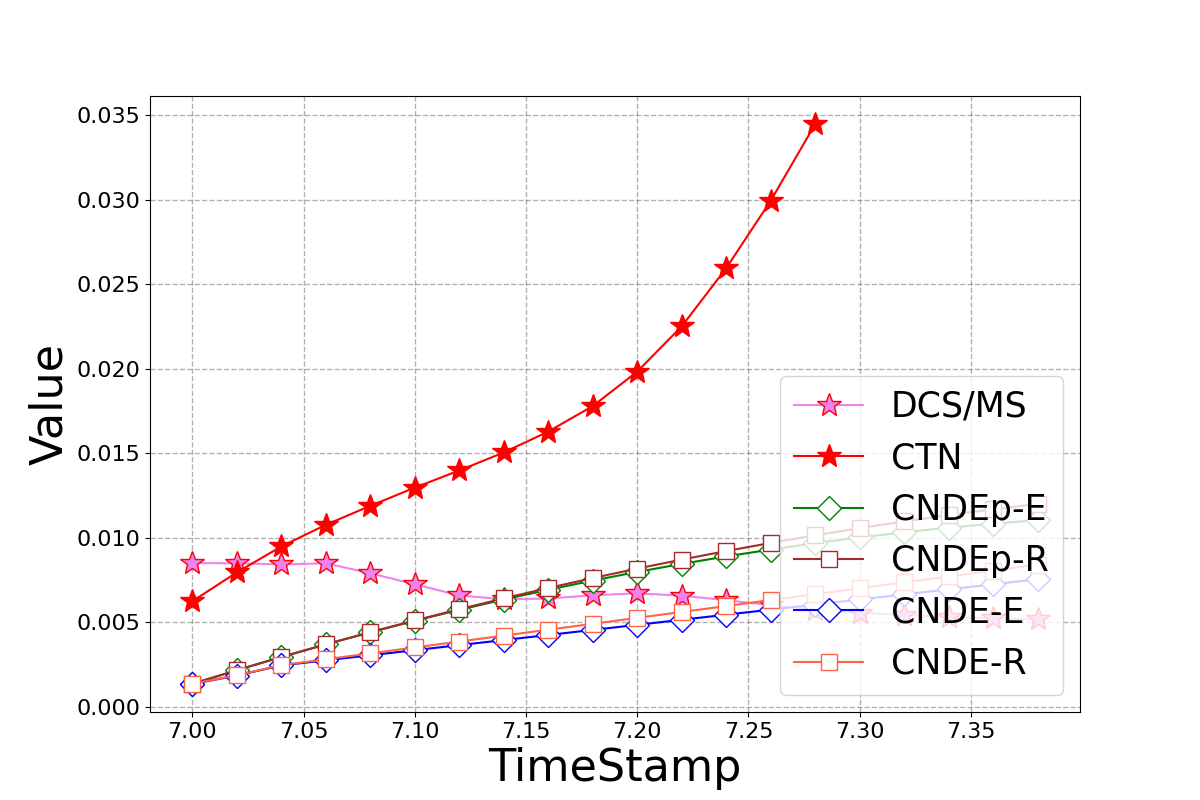}
}\hspace{-0.1in}
\subfigure[{$v$ Channel.}]{ \label{fig:b}{}
\includegraphics[width=0.32\linewidth]{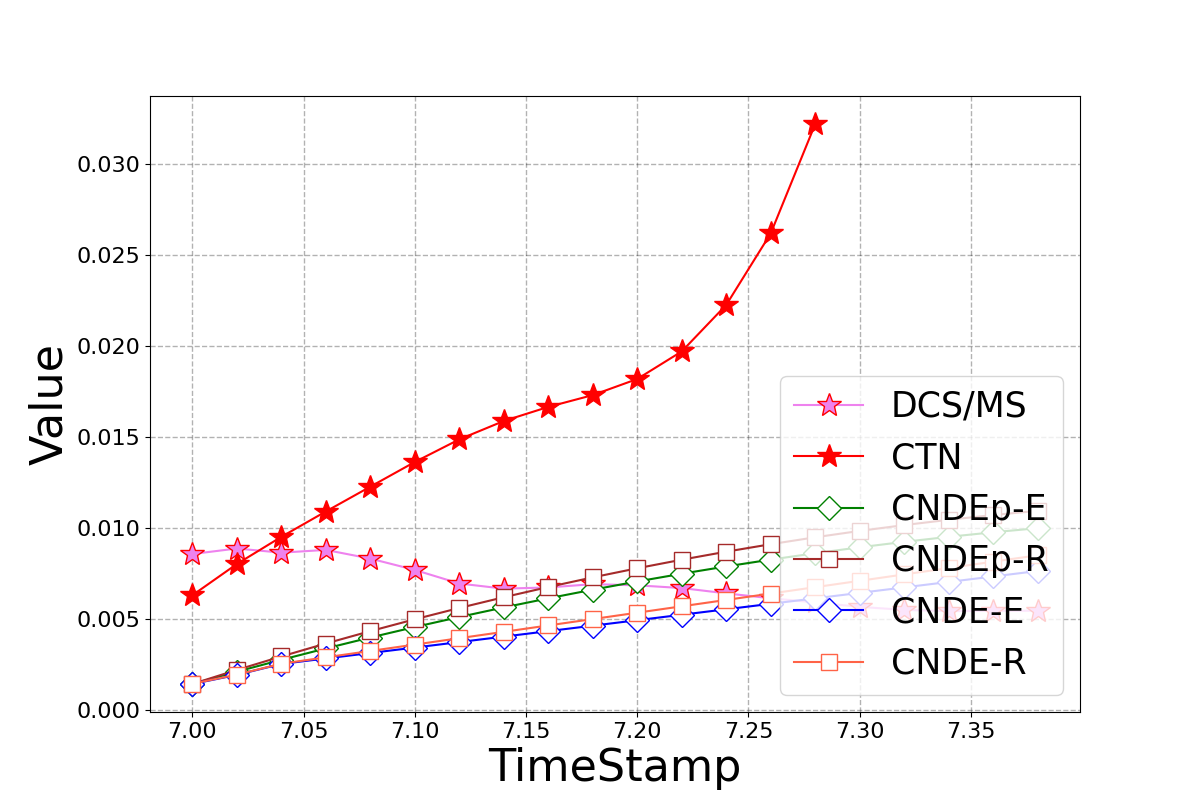}
}\hspace{-0.1in}
\subfigure[{$w$ Channel.}]{ \label{fig:b}{}
\includegraphics[width=0.32\linewidth]{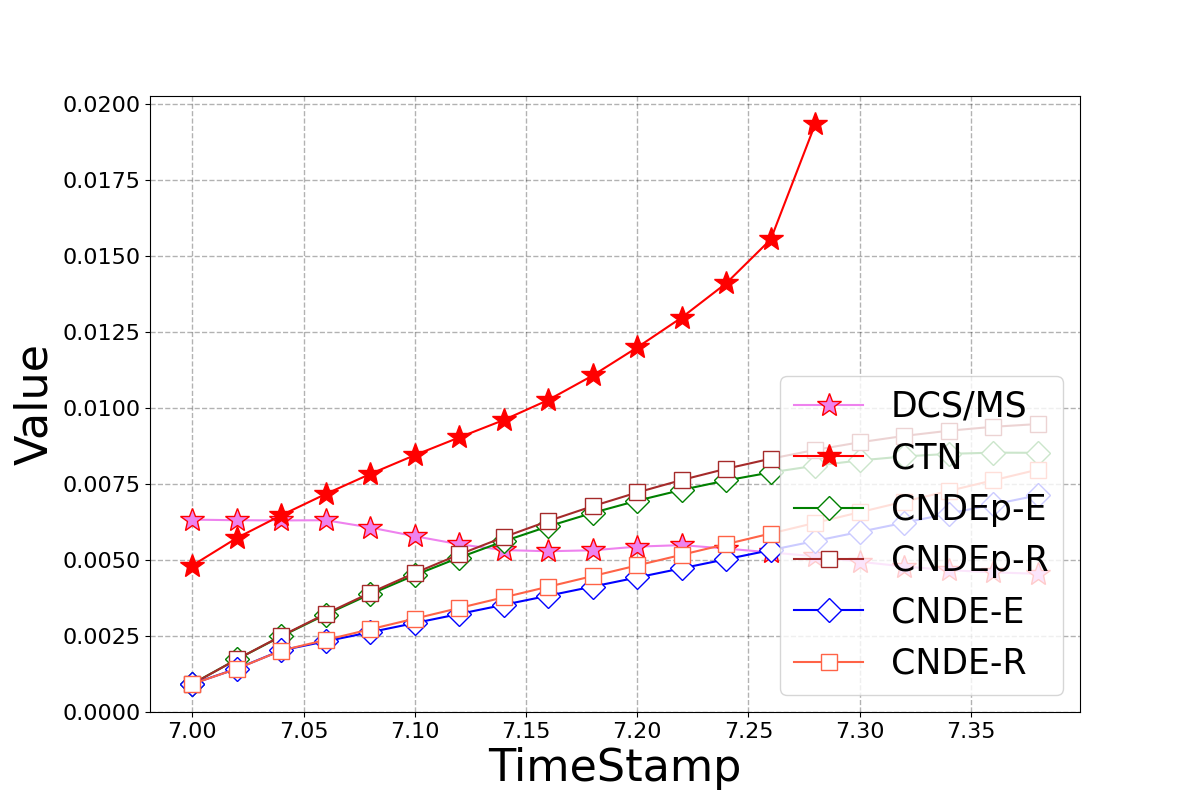}
}
\vspace{-.1in}
\caption{Change of dissipation difference produced by different models from 1st (80s) to 20th (120s) time step in the TGV dataset.}
\label{fig:tf_plot5_tgv}
\end{figure*}

\begin{figure*} [!h]
\centering
\subfigure[LES Upsampling.]{ \label{fig:a}
\includegraphics[width=0.16\linewidth]{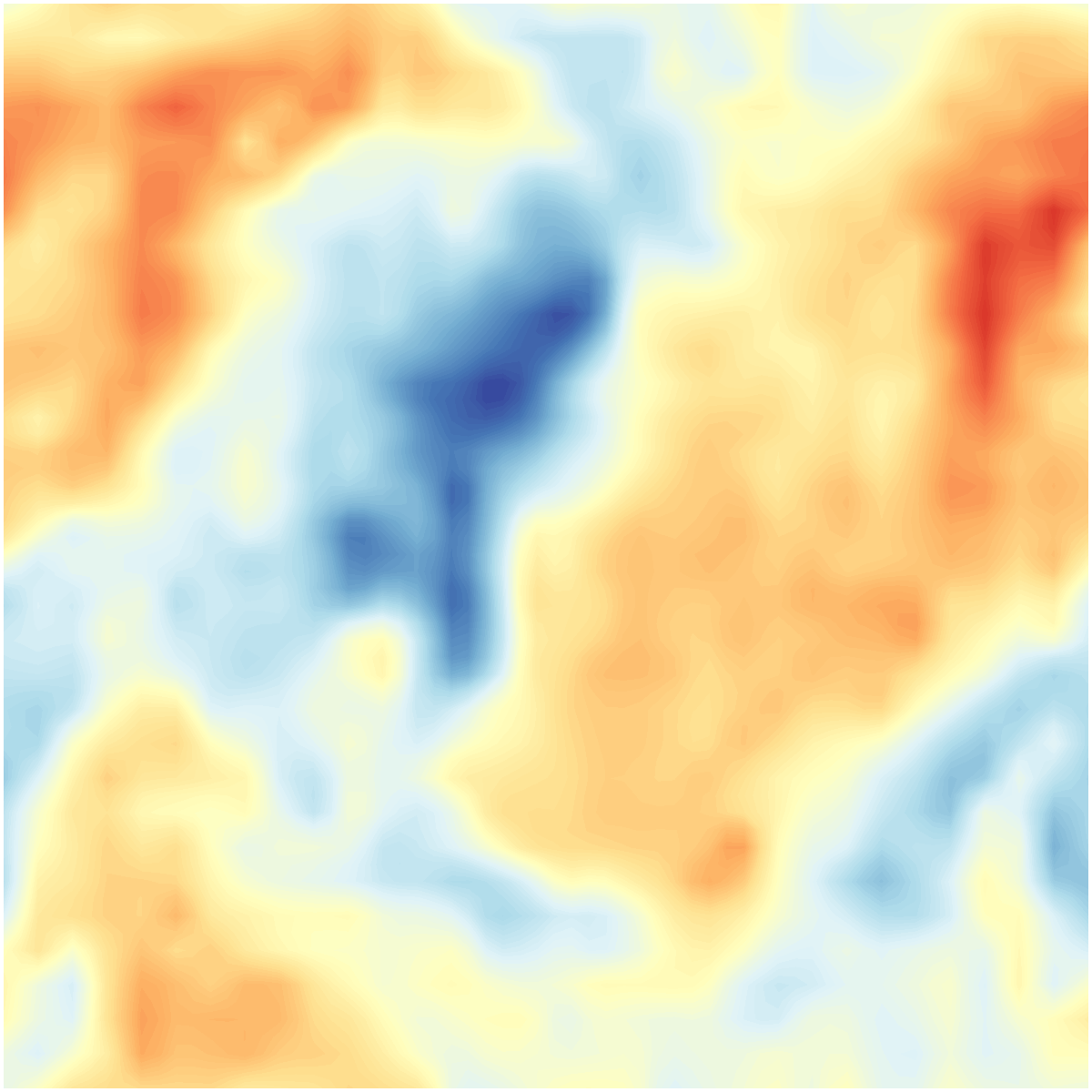}
}
\subfigure[DCS/MS.]{ \label{fig:b}
\includegraphics[width=0.16\linewidth]{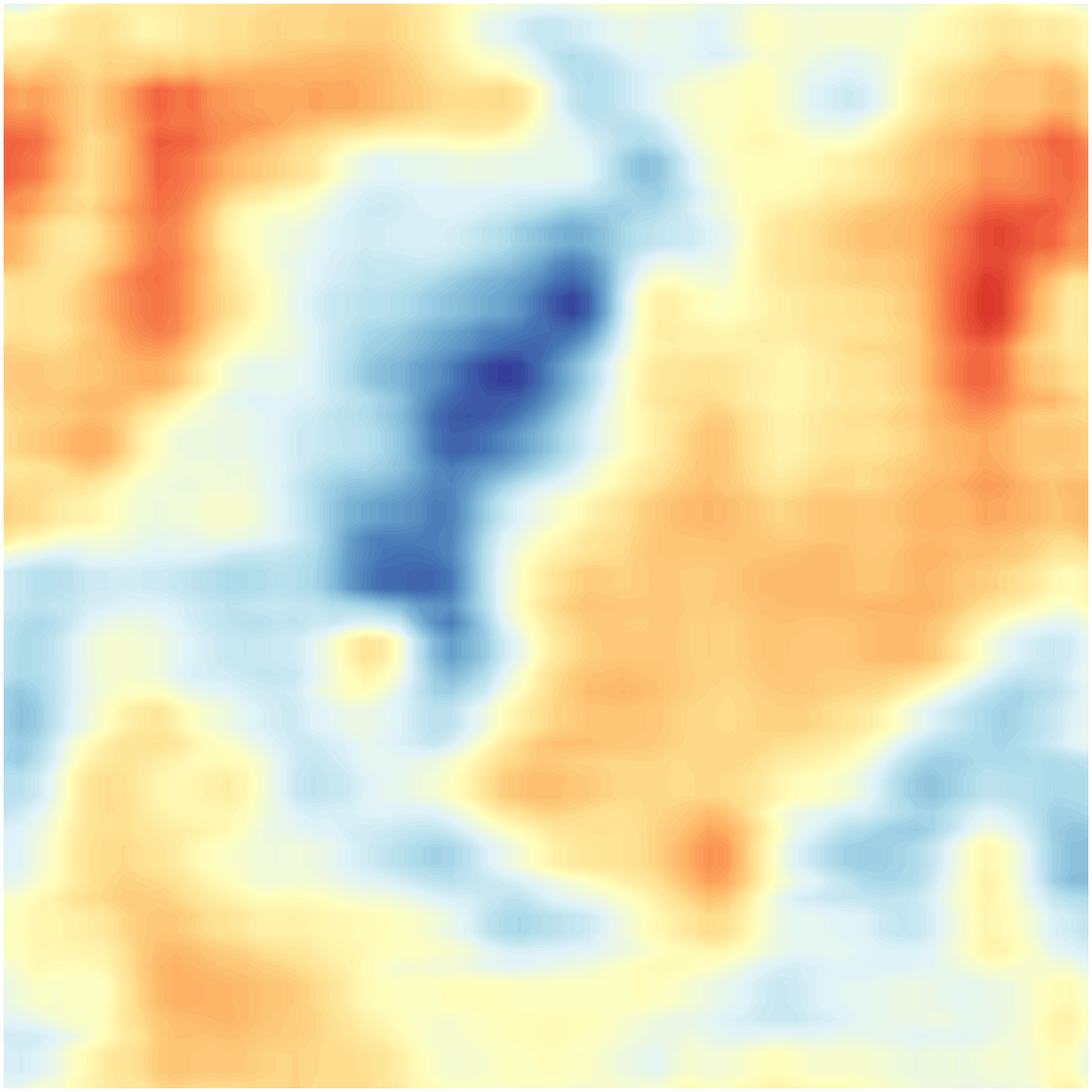}
}
\subfigure[CTN.]{ \label{fig:c}
\includegraphics[width=0.16\linewidth]{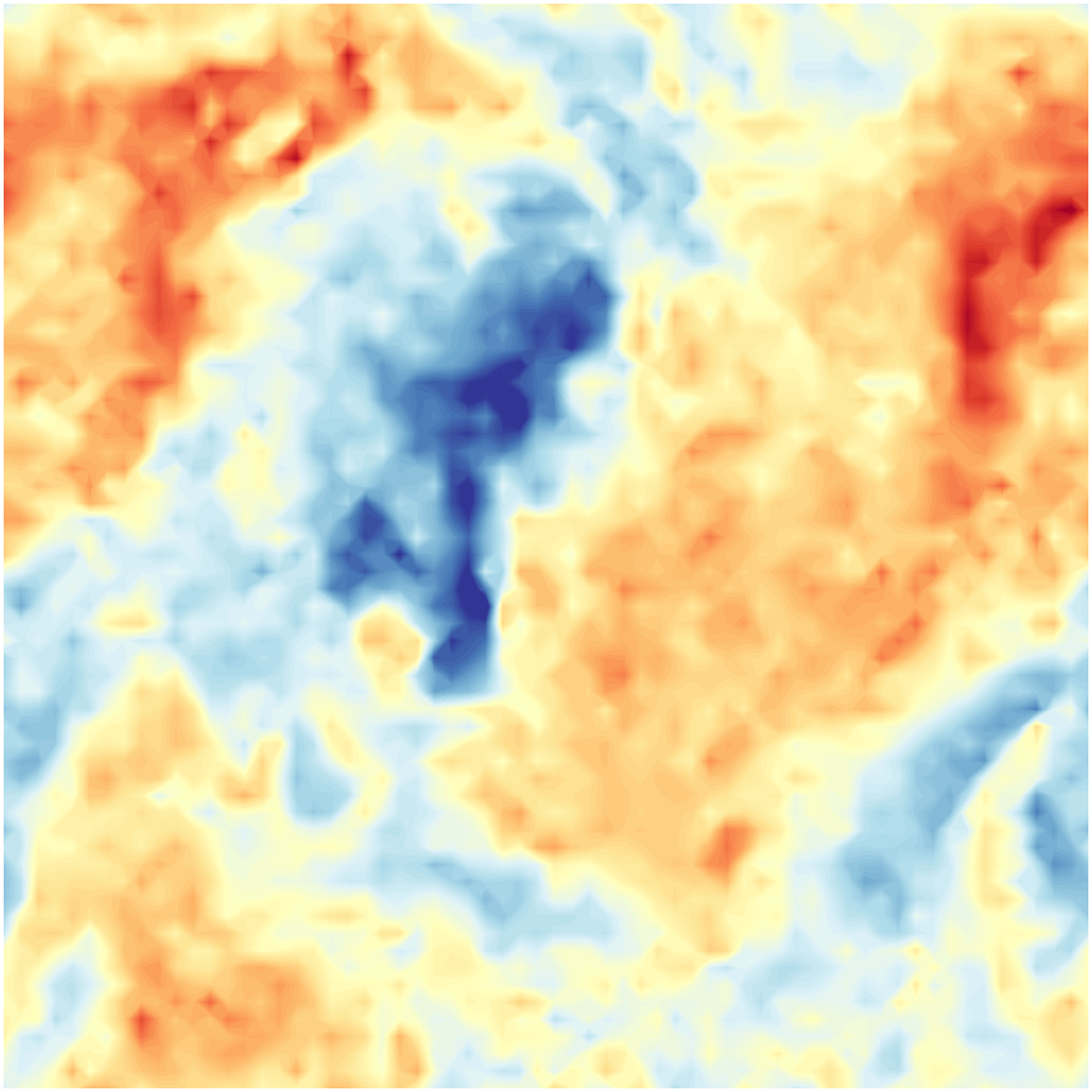}
}
\subfigure[CNDE-E.]{ \label{fig:d}
\includegraphics[width=0.16\linewidth]{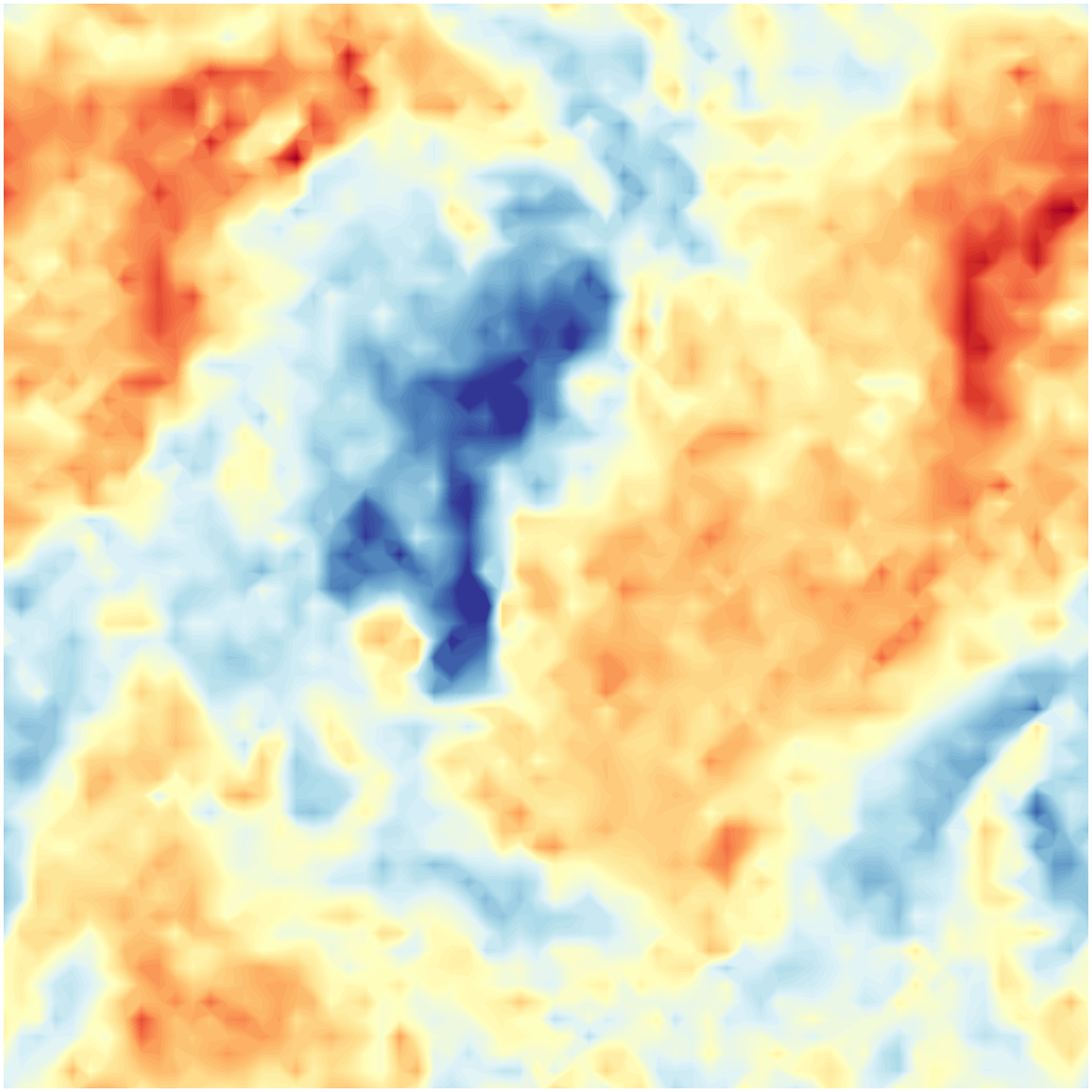}
}
\subfigure[CNDE-R.]{ \label{fig:e}
\includegraphics[width=0.16\linewidth]{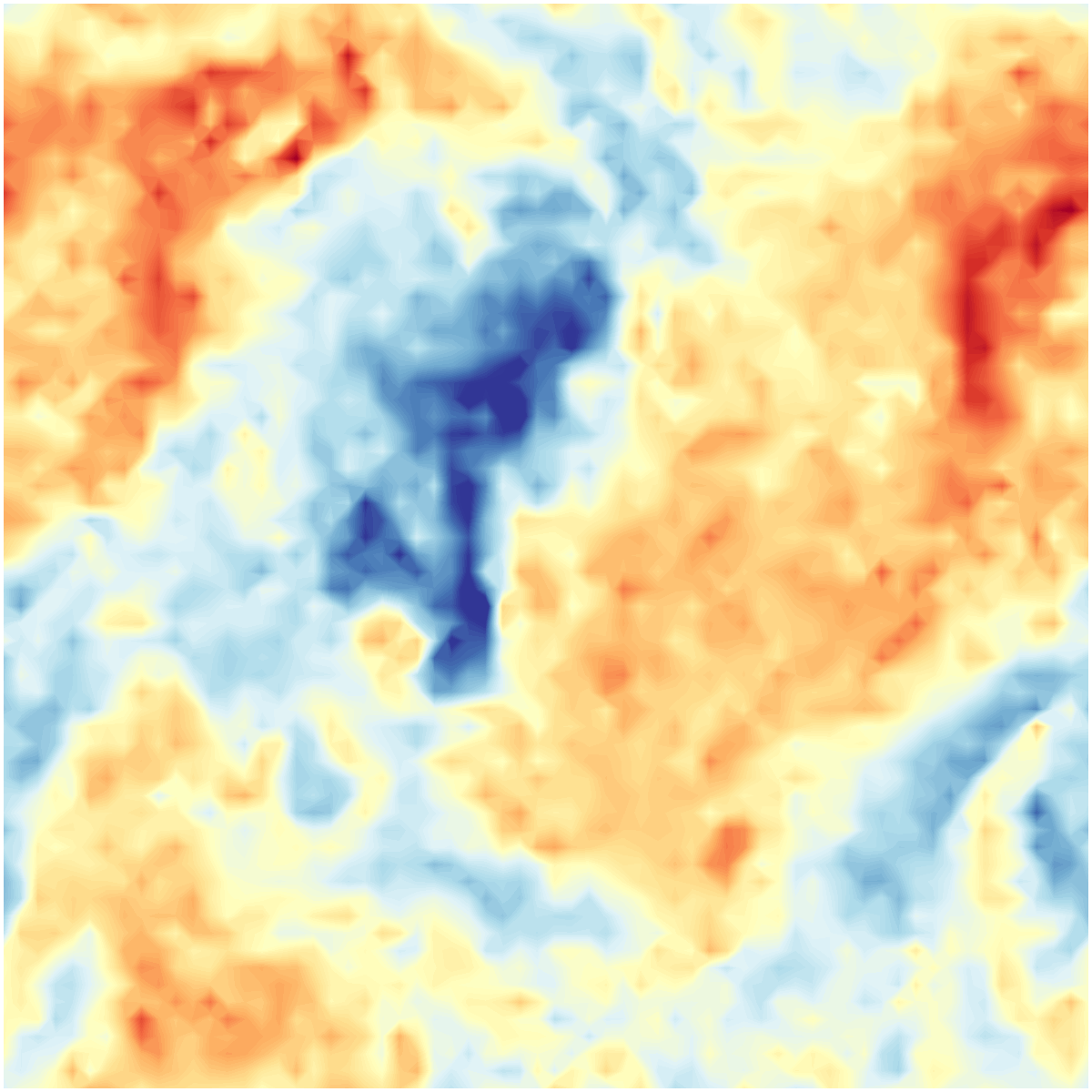}
}
\subfigure[Target DNS.]{ \label{fig:f}
\includegraphics[width=0.16\linewidth]{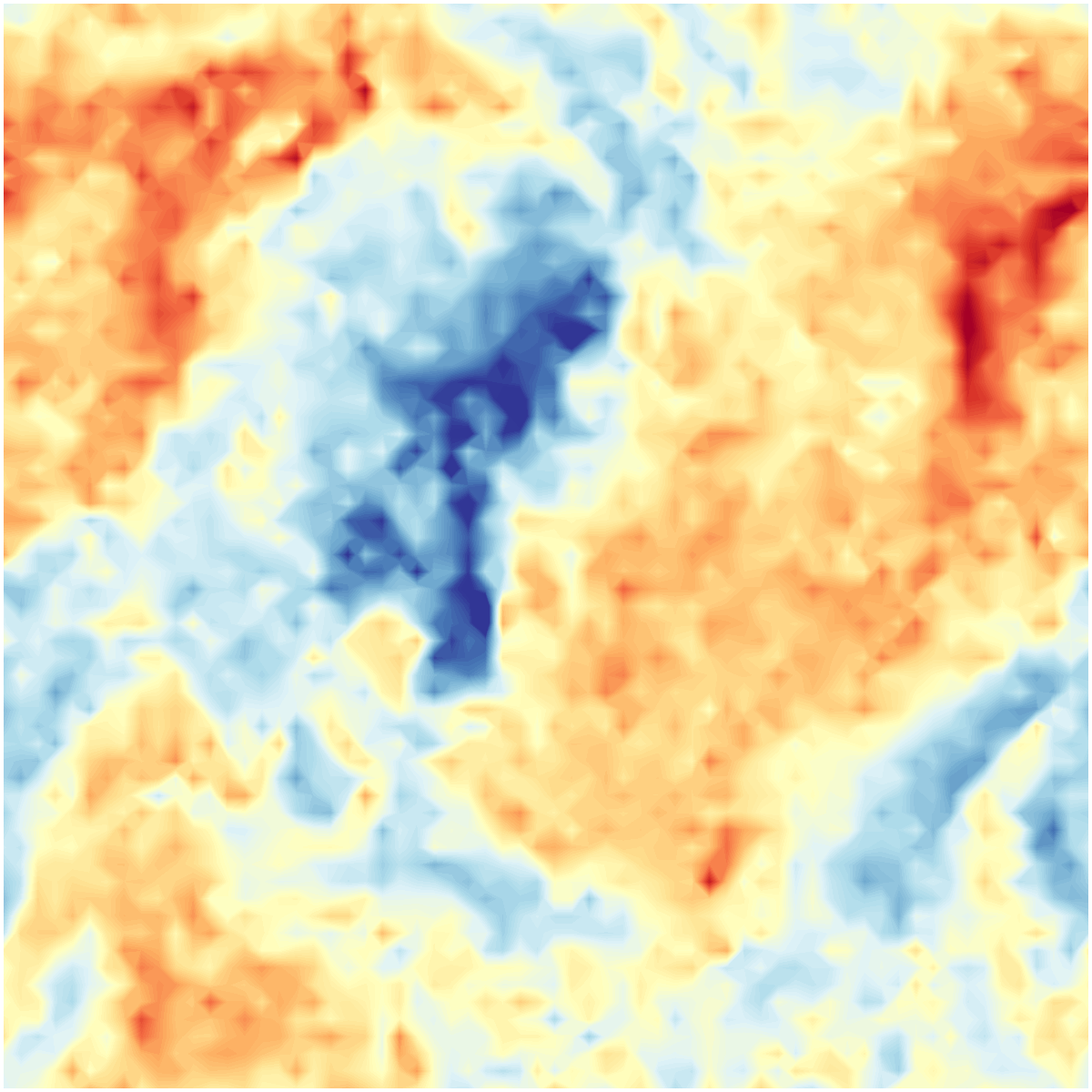}
}\vspace{-.1in}
\subfigure[LES Upsampling.]{ \label{fig:a}
\includegraphics[width=0.16\linewidth]{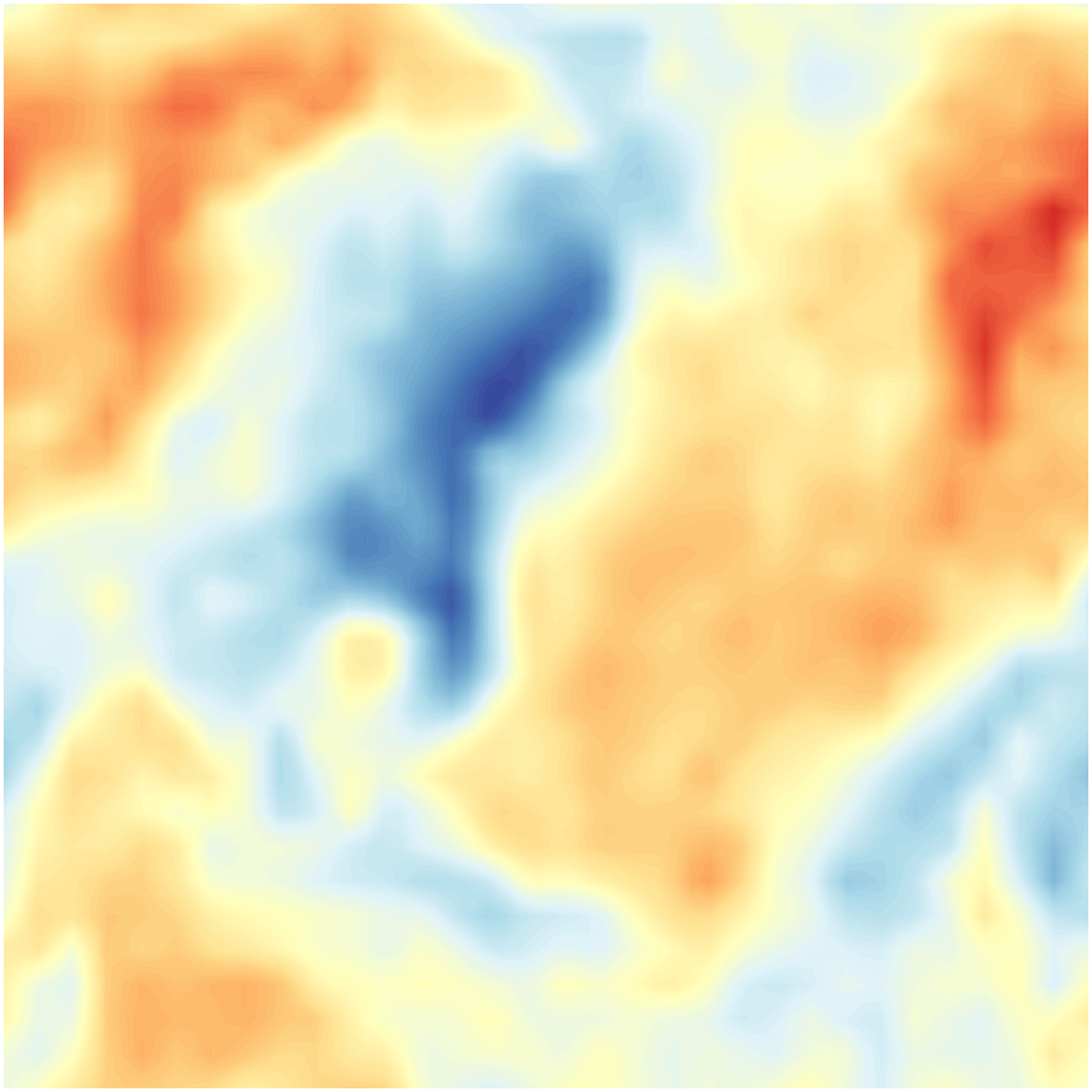}
}
\subfigure[DCS/MS.]{ \label{fig:b}
\includegraphics[width=0.16\linewidth]{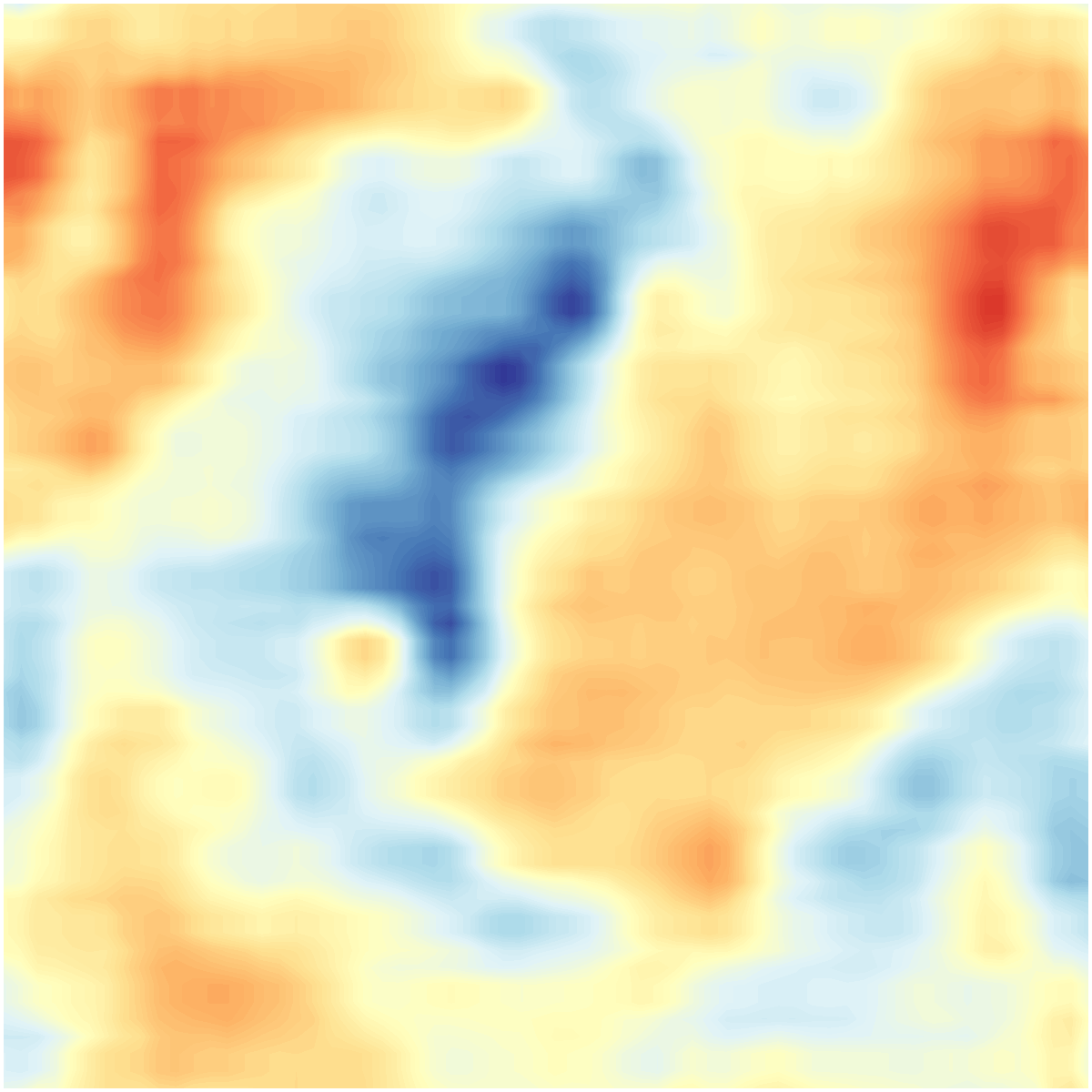}
}
\subfigure[CTN.]{ \label{fig:c}
\includegraphics[width=0.16\linewidth]{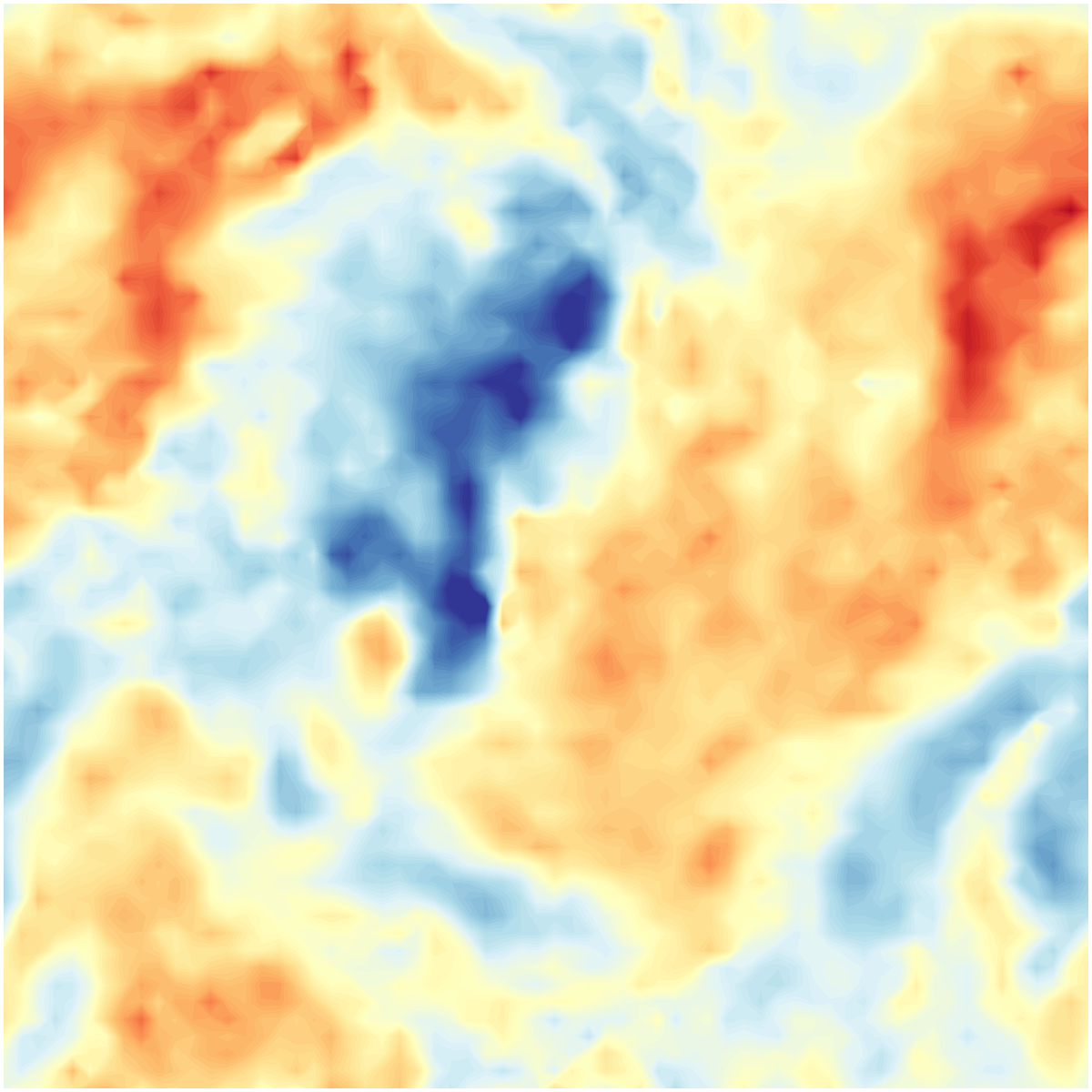}
}
\subfigure[{CNDE-E.}]{ \label{fig:d}
\includegraphics[width=0.16\linewidth]{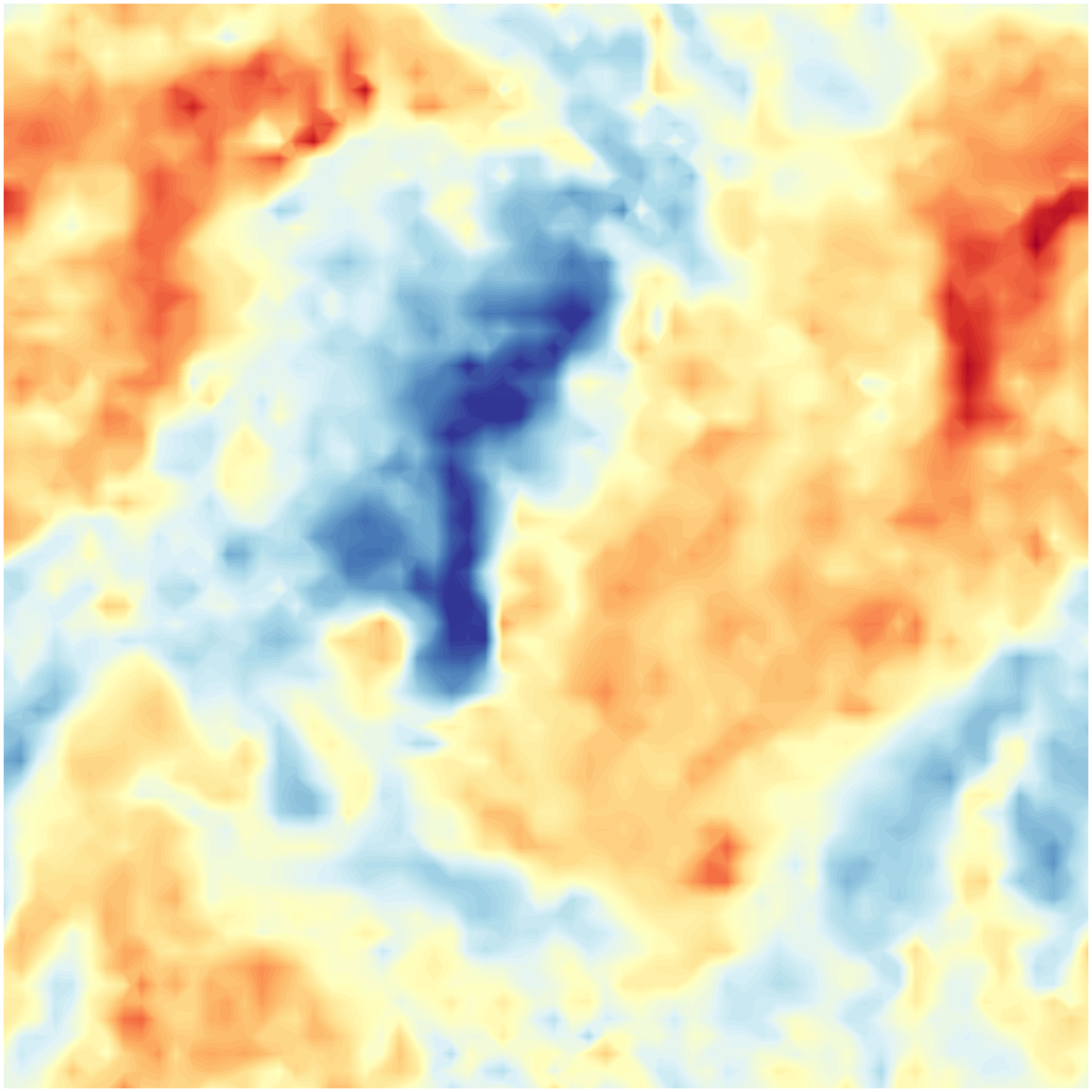}
}
\subfigure[{CNDE-R.}]{ \label{fig:e}
\includegraphics[width=0.16\linewidth]{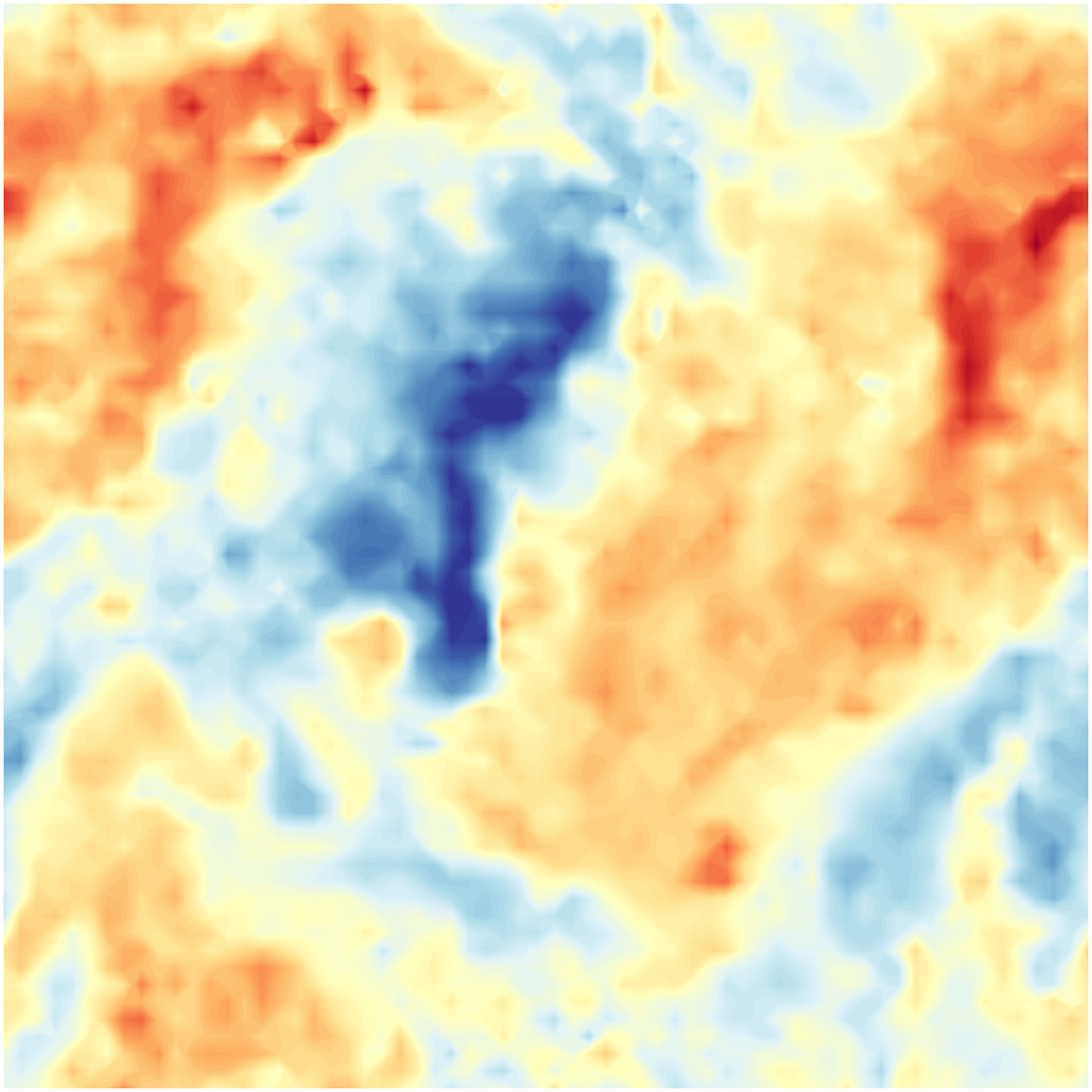}
}
\subfigure[Target DNS.]{ \label{fig:f}
\includegraphics[width=0.16\linewidth]{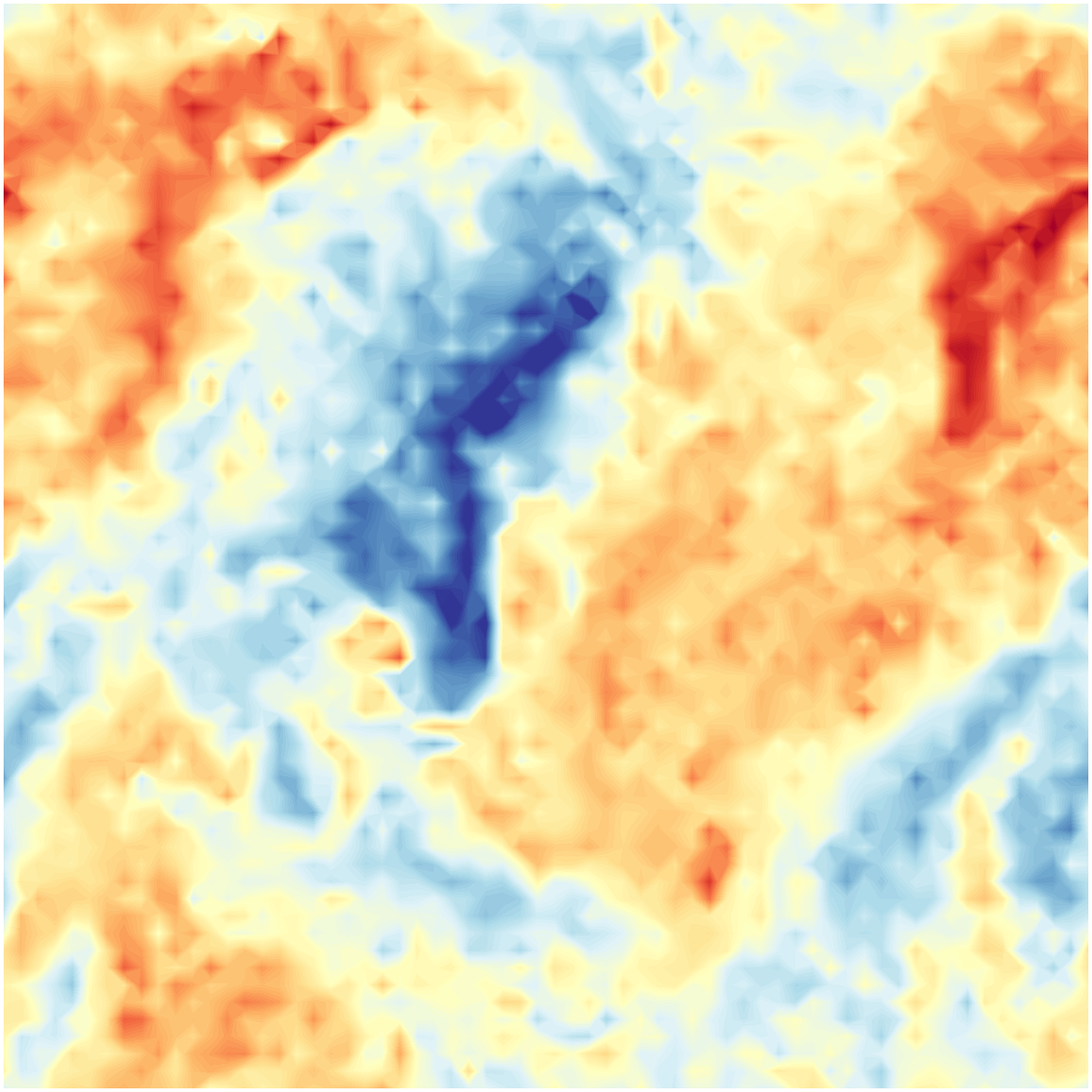}
}\vspace{-.1in}
\subfigure[LES Upsampling.]{ \label{fig:a}
\includegraphics[width=0.16\linewidth]{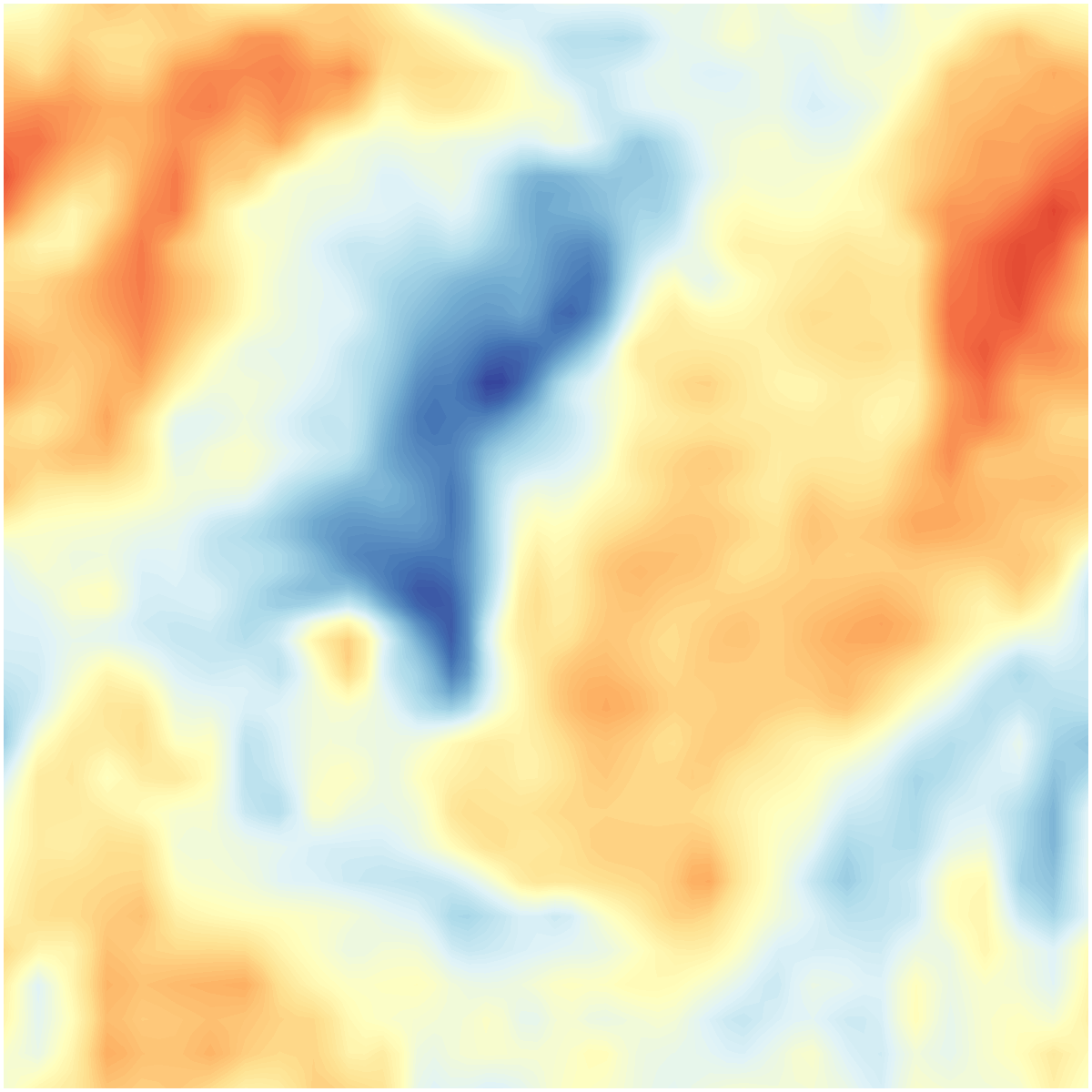}
}
\subfigure[DCS/MS.]{ \label{fig:b}
\includegraphics[width=0.16\linewidth]{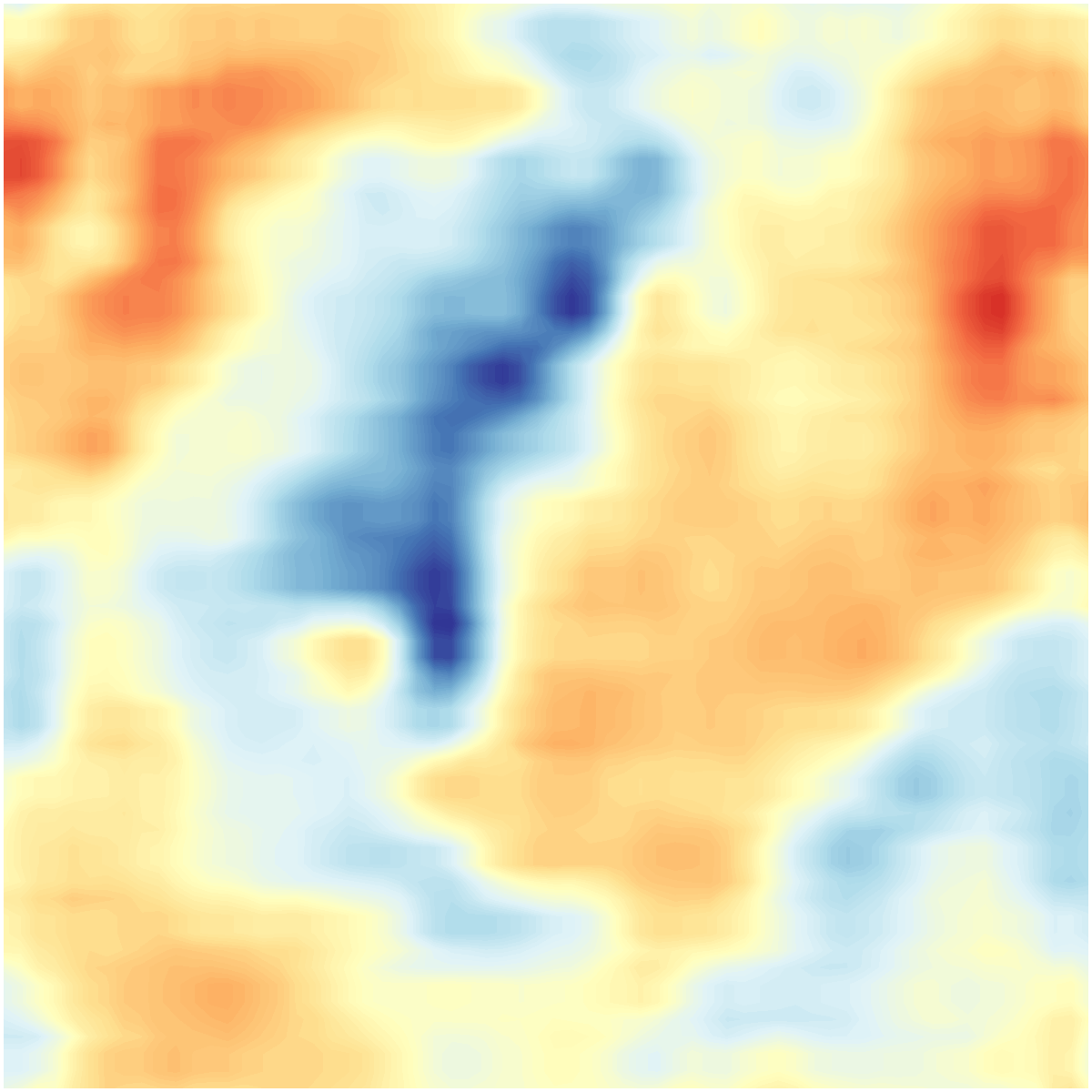}
}
\subfigure[CTN.]{ \label{fig:c}
\includegraphics[width=0.16\linewidth]{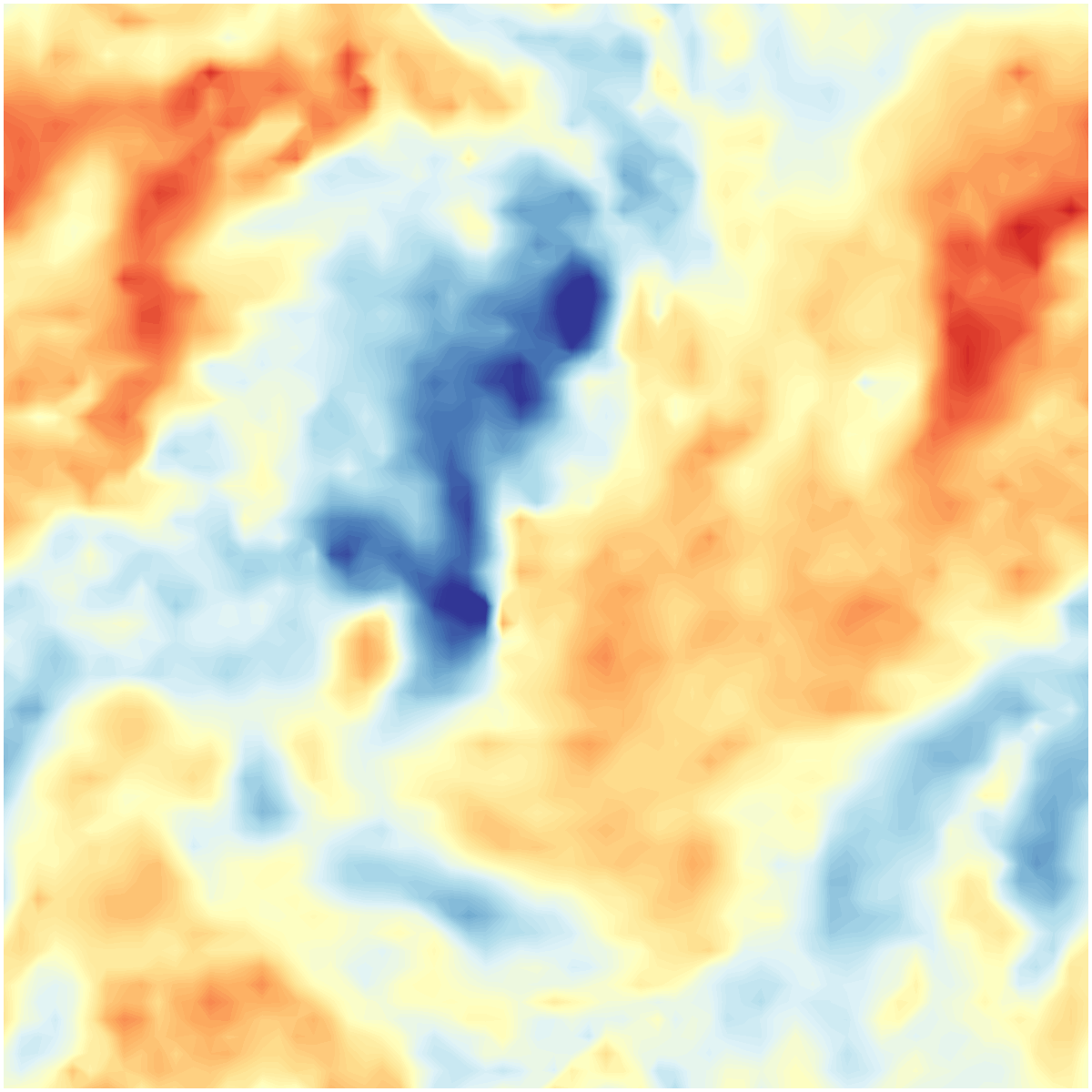}
}
\subfigure[{CNDE-E.}]{ \label{fig:d}
\includegraphics[width=0.16\linewidth]{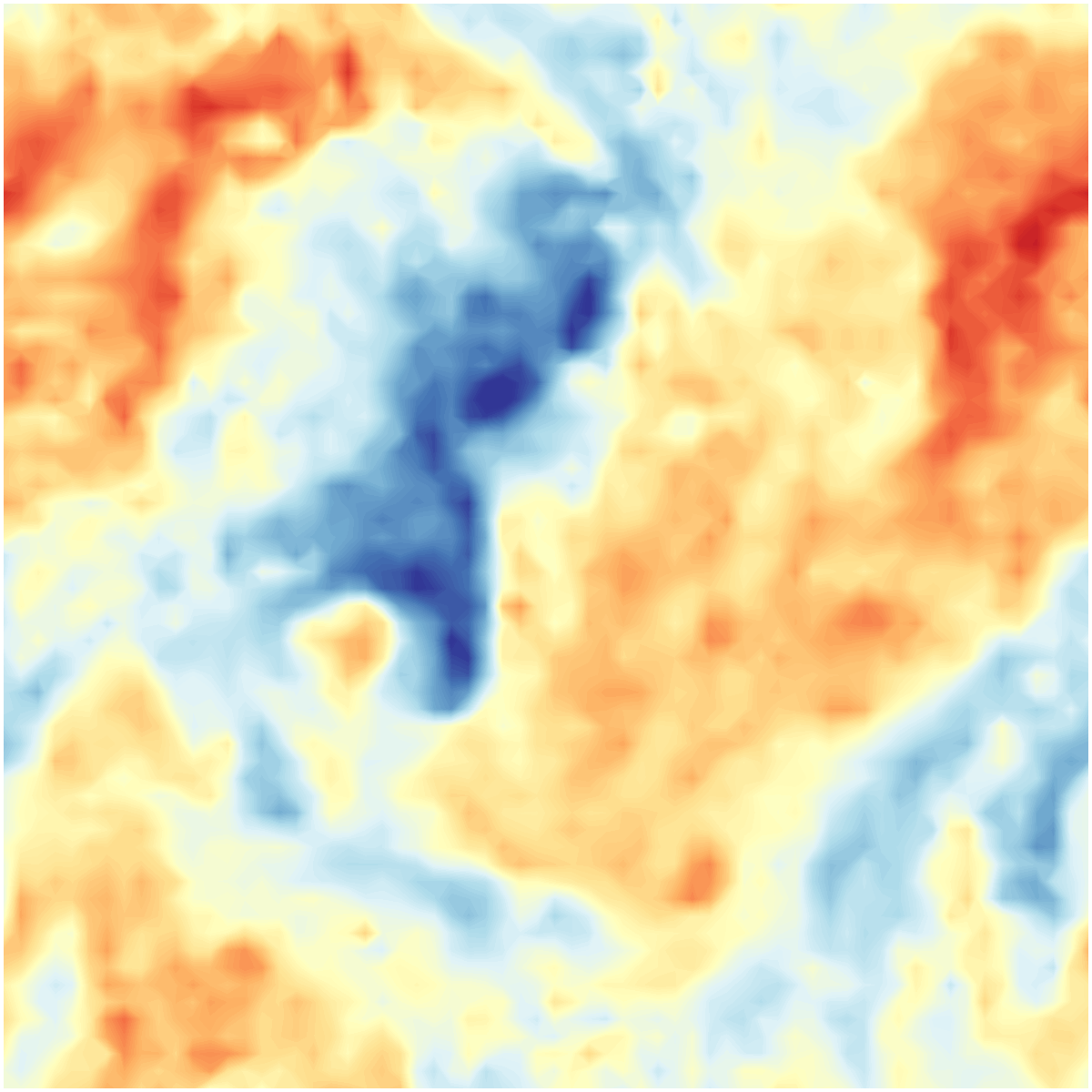}
}
\subfigure[{CNDE-R.}]{ \label{fig:e}
\includegraphics[width=0.16\linewidth]{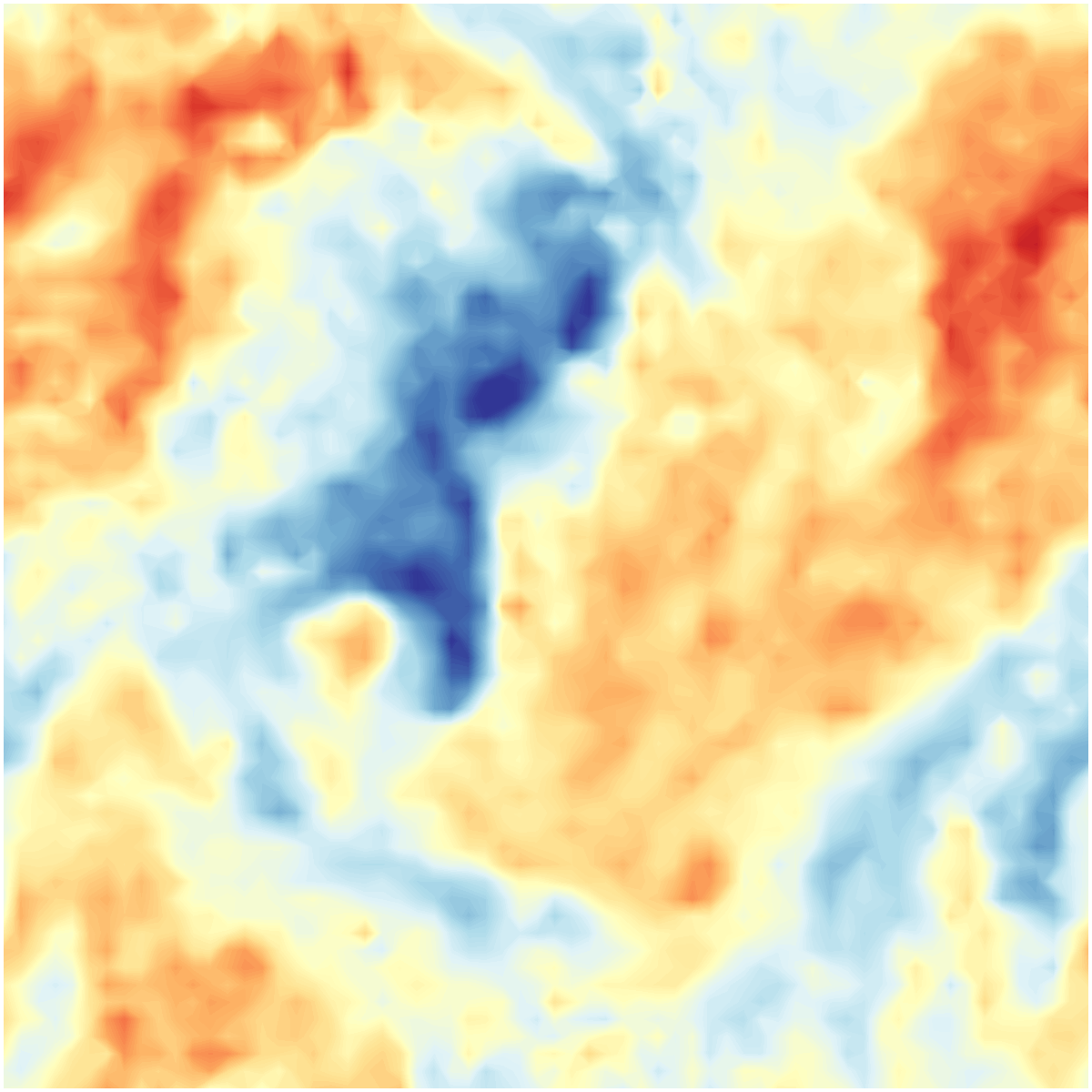}
}
\subfigure[Target DNS.]{ \label{fig:f}
\includegraphics[width=0.16\linewidth]{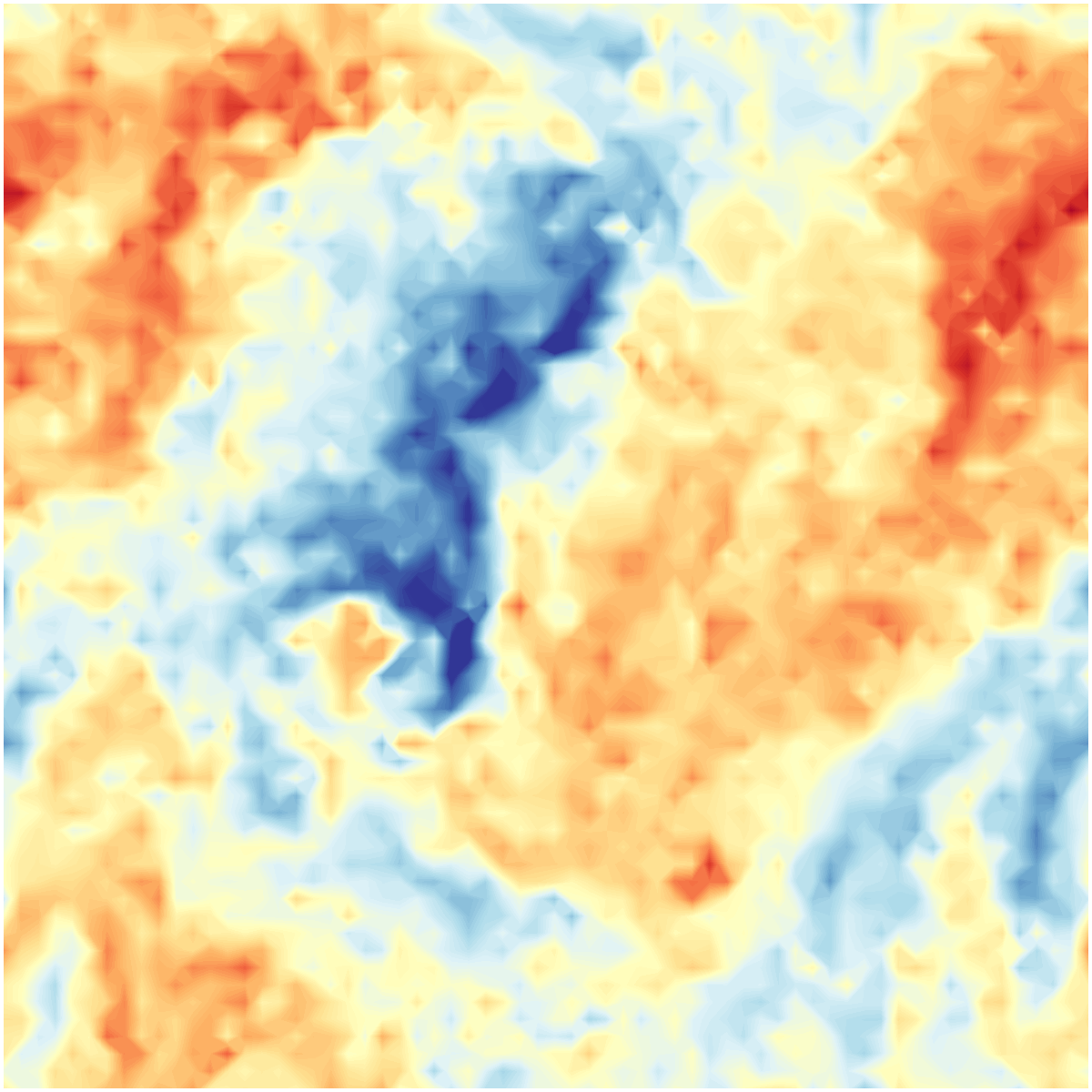}
}\vspace{-.1in}
\subfigure[LES Upsampling.]{ \label{fig:a}
\includegraphics[width=0.16\linewidth]{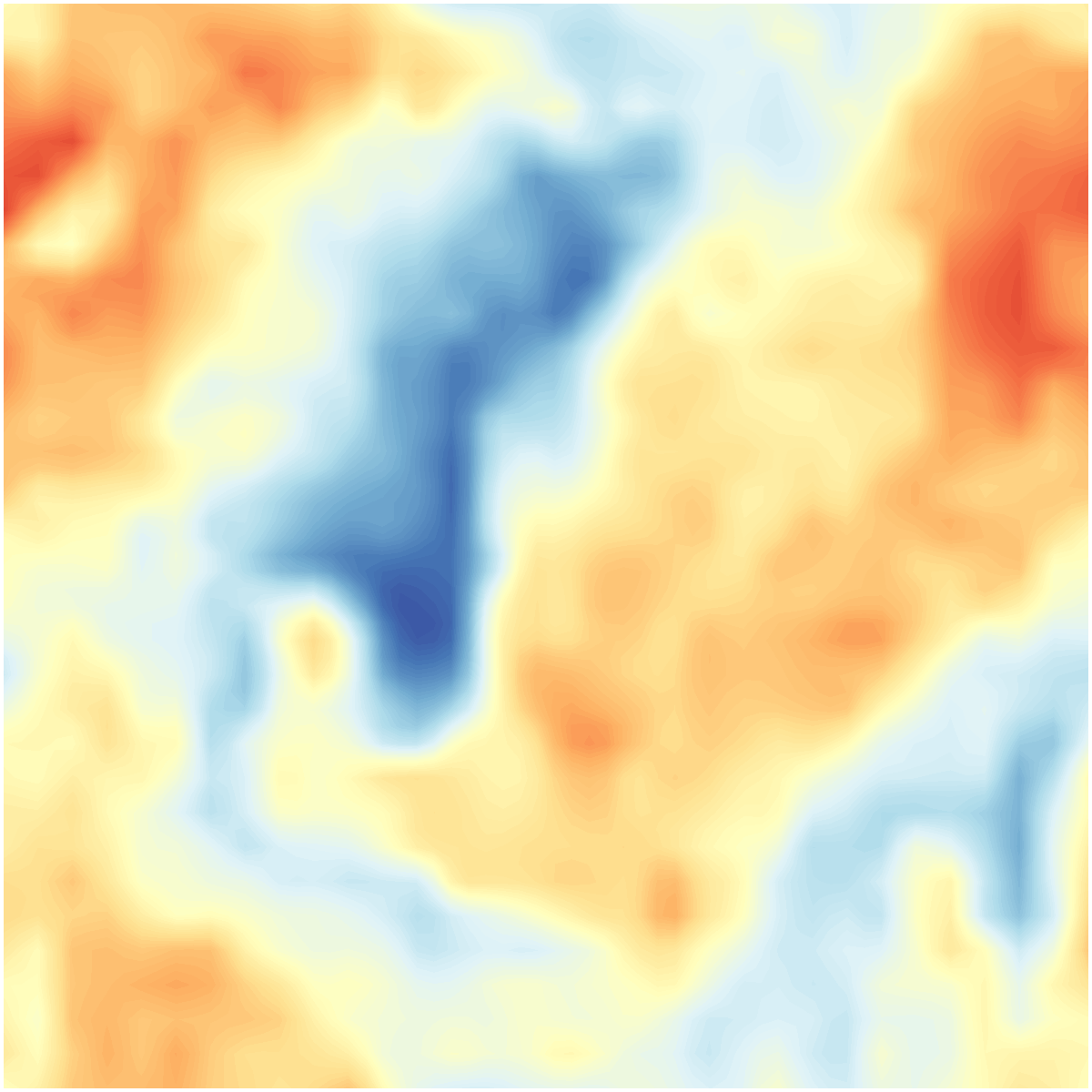}
}
\subfigure[DCS/MS.]{ \label{fig:b}
\includegraphics[width=0.16\linewidth]{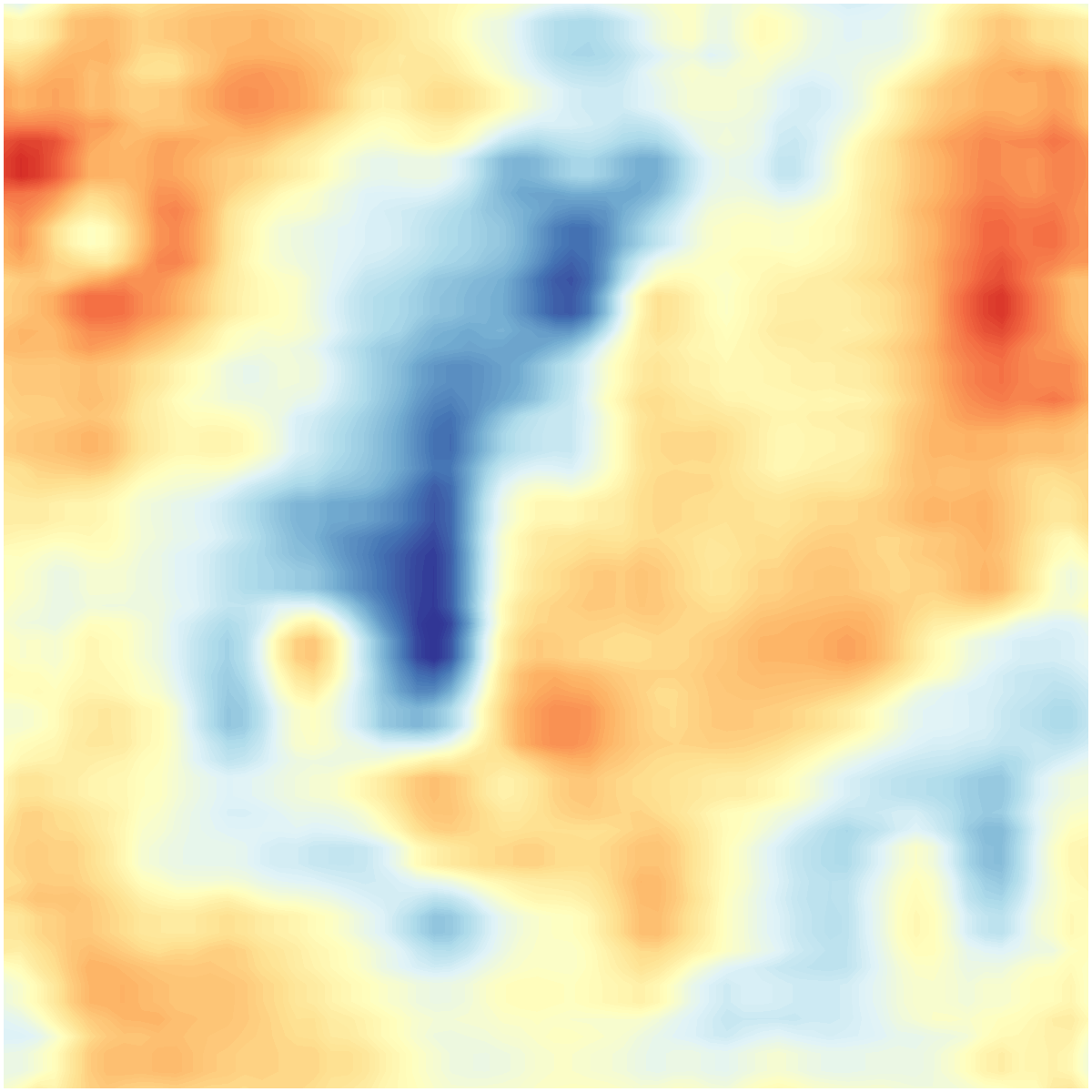}
}
\subfigure[CTN.]{ \label{fig:c}
\includegraphics[width=0.16\linewidth]{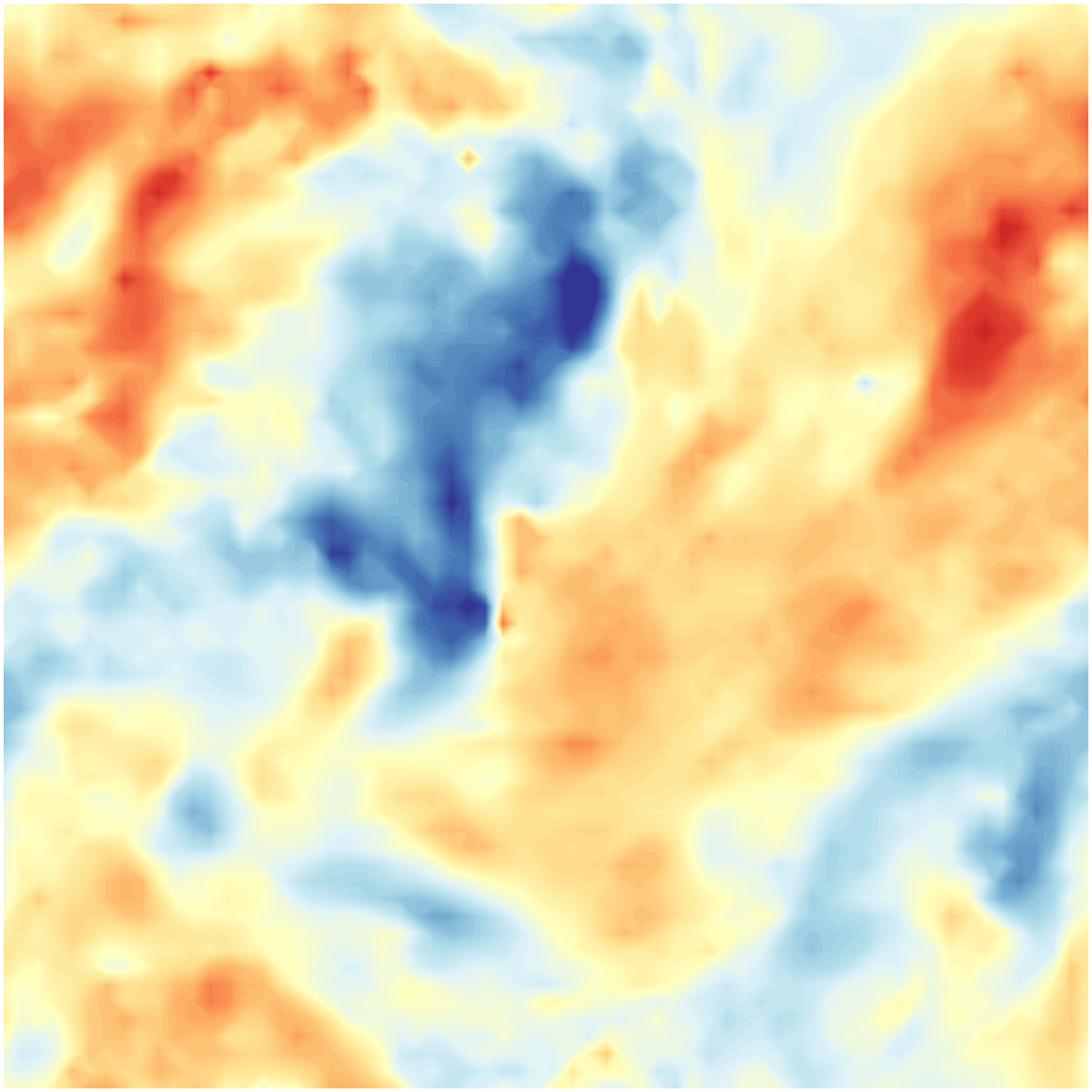}
}
\subfigure[{CNDE-E.}]{ \label{fig:d}
\includegraphics[width=0.16\linewidth]{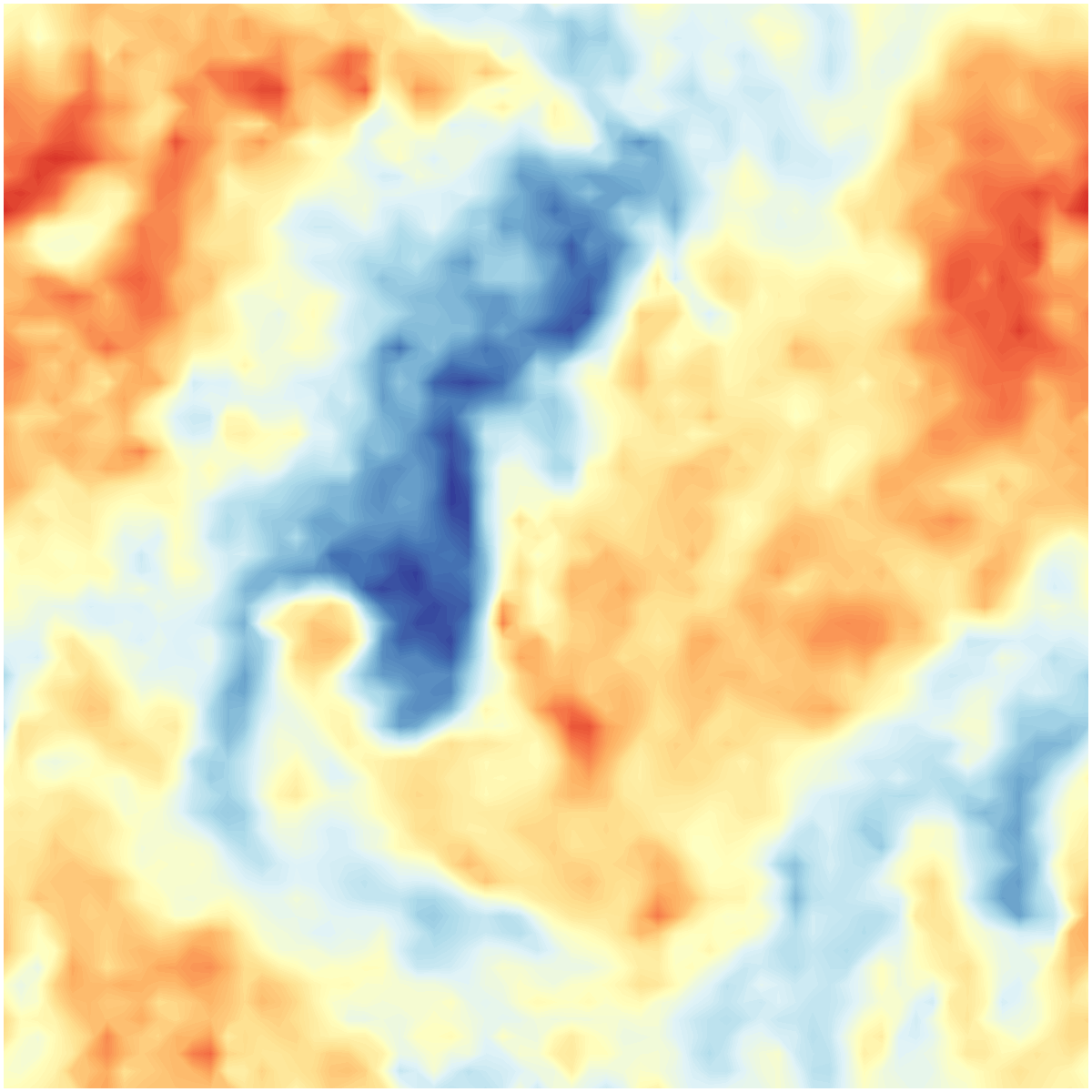}
}
\subfigure[{CNDE-R.}]{ \label{fig:e}
\includegraphics[width=0.16\linewidth]{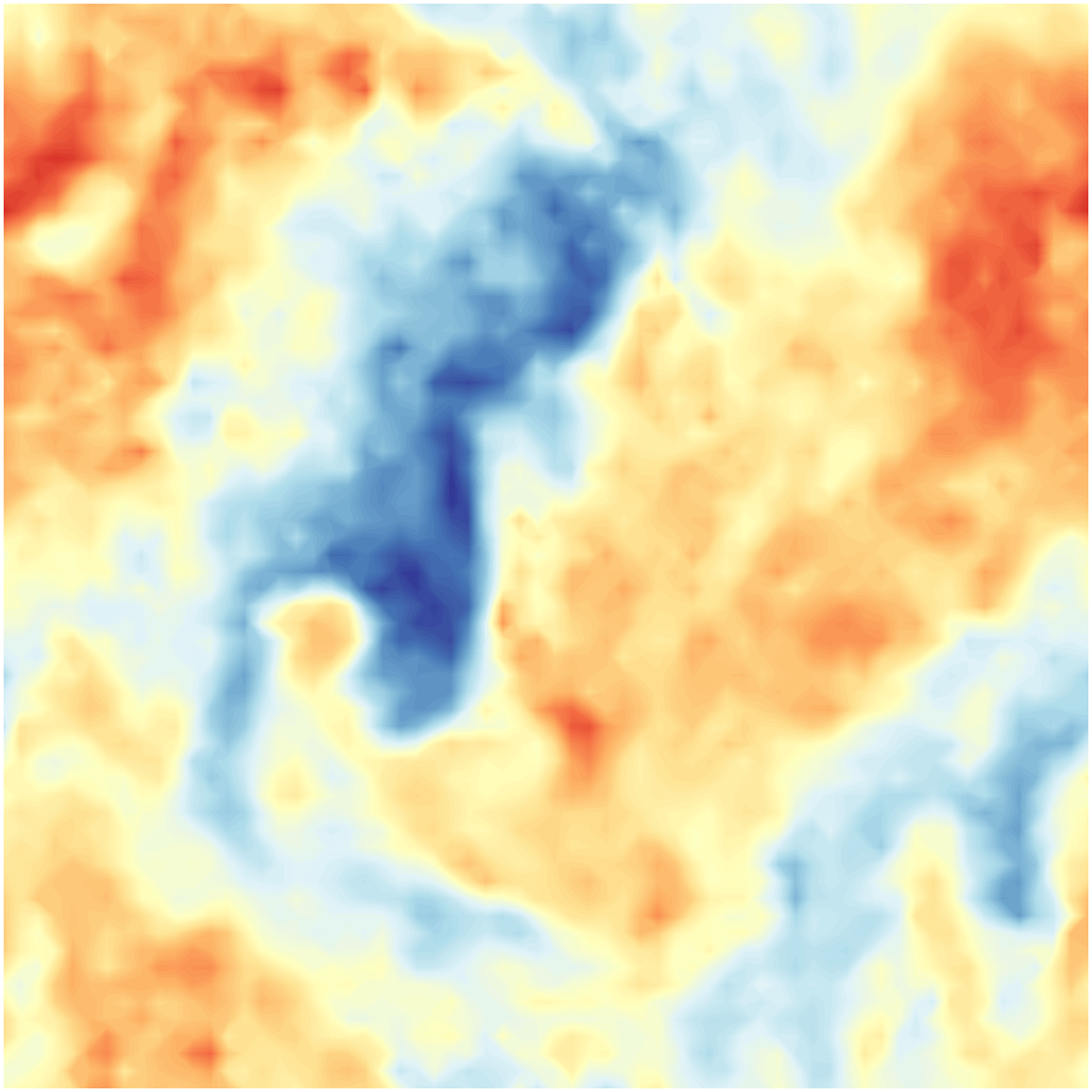}
}
\subfigure[Target DNS.]{ \label{fig:f}
\includegraphics[width=0.16\linewidth]{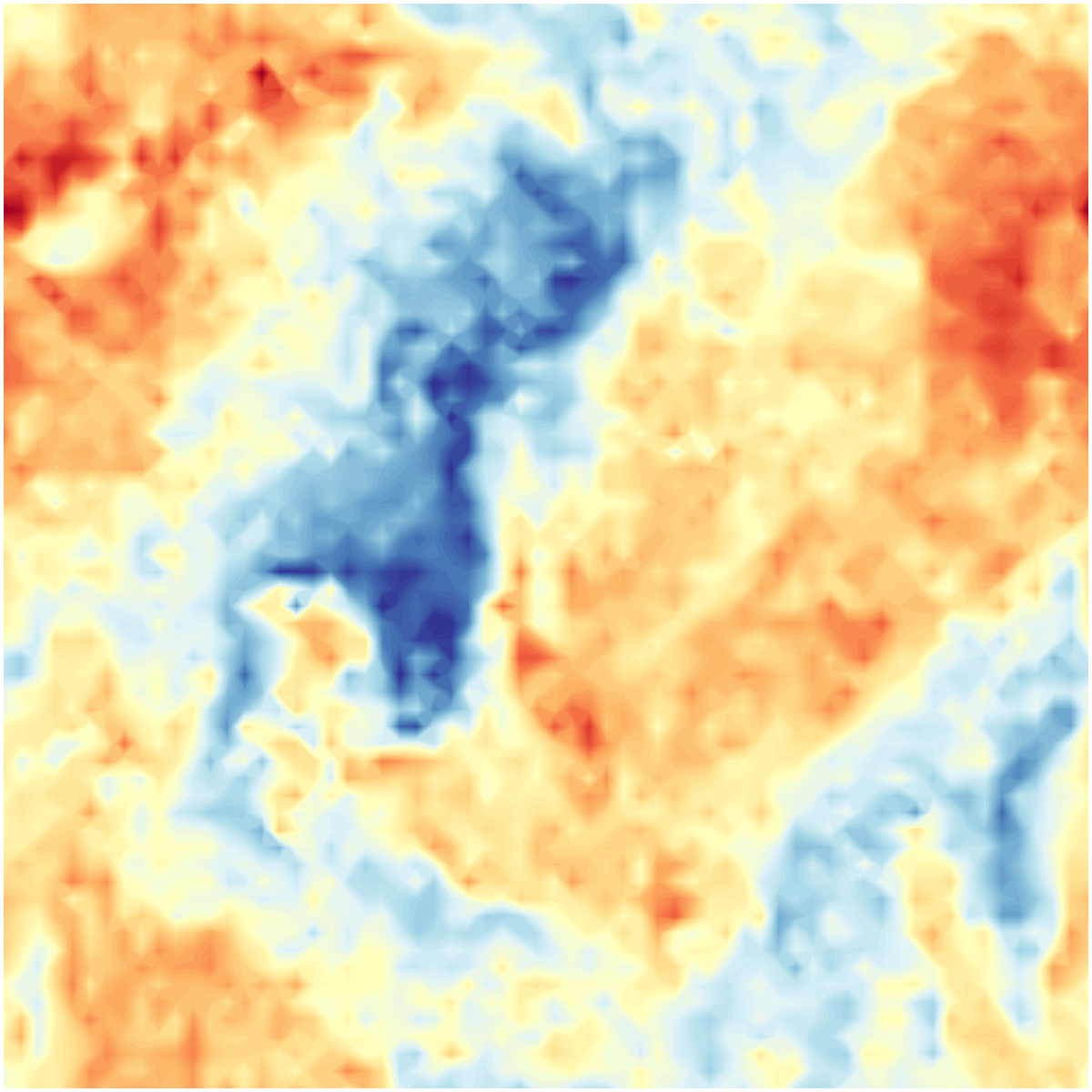}
}
\caption{Reconstructed $w$ channel by each method on a sample testing slice along the $z$ dimension in the FIT dataset. The reconstruction results are shown at 1st (7s), 5th (7.1s) 10th (7.2s) and 20th (7.4s) in (a)-(f), (g)-(l), (m)-(r) and (s)-(x), respectively.}
\label{fig:tf_plot3}
\end{figure*}

\begin{figure*} [!h]
\centering
\subfigure[LES Upsampling.]{ \label{fig:a}
\includegraphics[width=0.16\linewidth]{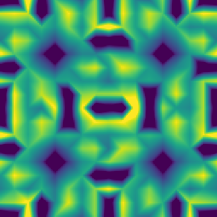}
}
\subfigure[DCS/MS.]{ \label{fig:b}
\includegraphics[width=0.16\linewidth]{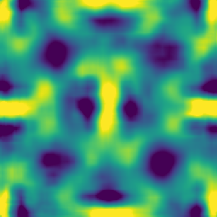}
}
\subfigure[CTN.]{ \label{fig:c}
\includegraphics[width=0.16\linewidth]{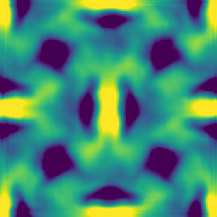}
}
\subfigure[CNDE-E.]{ \label{fig:d}
\includegraphics[width=0.16\linewidth]{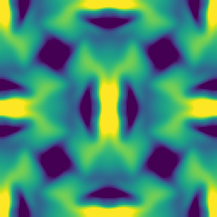}
}
\subfigure[CNDE-R.]{ \label{fig:e}
\includegraphics[width=0.16\linewidth]{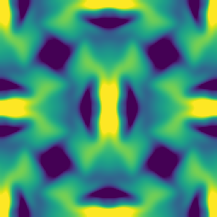}
}
\subfigure[Target DNS.]{ \label{fig:f}
\includegraphics[width=0.16\linewidth]{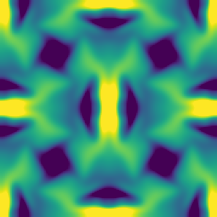}
}\vspace{-.1in}
\subfigure[LES Upsampling.]{ \label{fig:a}
\includegraphics[width=0.16\linewidth]{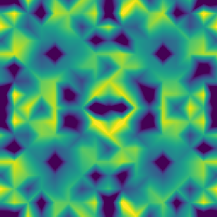}
}
\subfigure[DCS/MS.]{ \label{fig:b}
\includegraphics[width=0.16\linewidth]{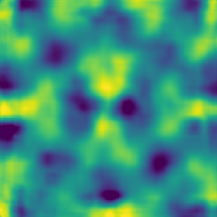}
}
\subfigure[CTN.]{ \label{fig:c}
\includegraphics[width=0.16\linewidth]{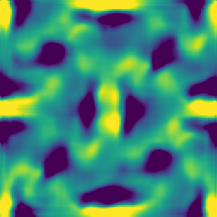}
}
\subfigure[{CNDE-E.}]{ \label{fig:d}
\includegraphics[width=0.16\linewidth]{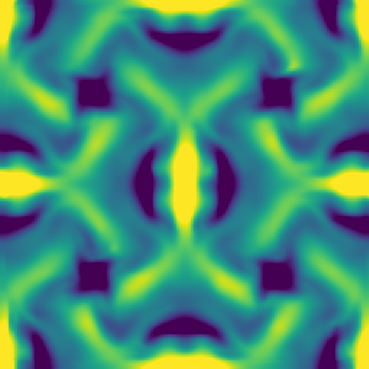}
}
\subfigure[{CNDE-R.}]{ \label{fig:e}
\includegraphics[width=0.16\linewidth]{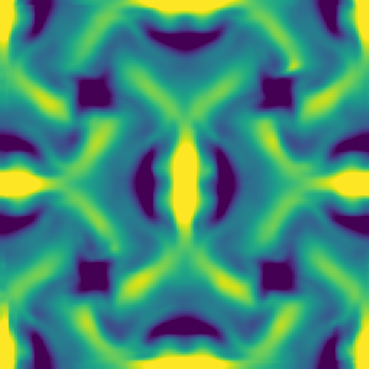}
}
\subfigure[Target DNS.]{ \label{fig:f}
\includegraphics[width=0.16\linewidth]{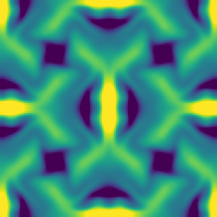}
}\vspace{-.1in}
\subfigure[LES Upsampling.]{ \label{fig:a}
\includegraphics[width=0.16\linewidth]{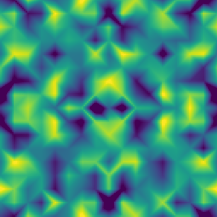}
}
\subfigure[DCS/MS.]{ \label{fig:b}
\includegraphics[width=0.16\linewidth]{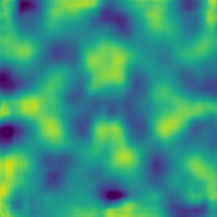}
}
\subfigure[CTN.]{ \label{fig:c}
\includegraphics[width=0.16\linewidth]{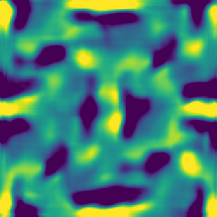}
}
\subfigure[{CNDE-E.}]{ \label{fig:d}
\includegraphics[width=0.16\linewidth]{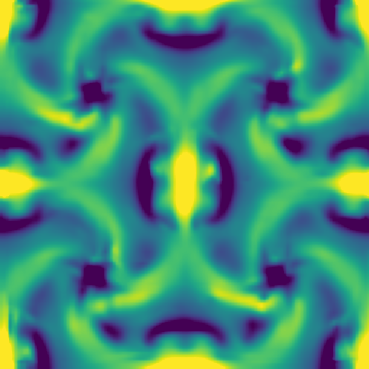}
}
\subfigure[{CNDE-R.}]{ \label{fig:e}
\includegraphics[width=0.16\linewidth]{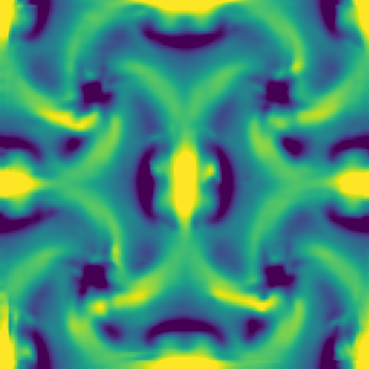}
}
\subfigure[Target DNS.]{ \label{fig:f}
\includegraphics[width=0.16\linewidth]{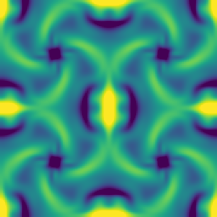}
}\vspace{-.1in}
\subfigure[LES Upsampling.]{ \label{fig:a}
\includegraphics[width=0.16\linewidth]{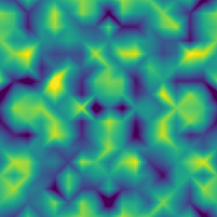}
}
\subfigure[DCS/MS.]{ \label{fig:b}
\includegraphics[width=0.16\linewidth]{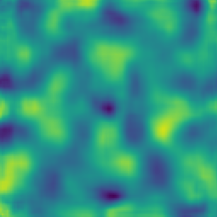}
}
\subfigure[CTN.]{ \label{fig:c}
\includegraphics[width=0.16\linewidth]{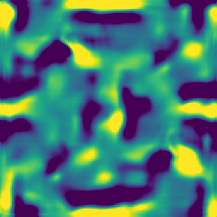}
}
\subfigure[{CNDE-E.}]{ \label{fig:d}
\includegraphics[width=0.16\linewidth]{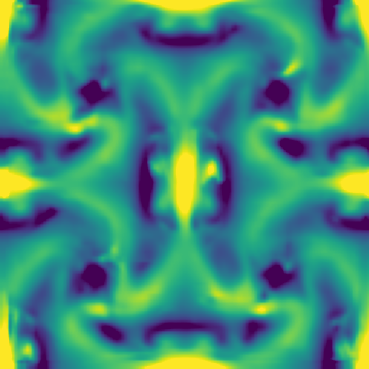}
}
\subfigure[{CNDE-R.}]{ \label{fig:e}
\includegraphics[width=0.16\linewidth]{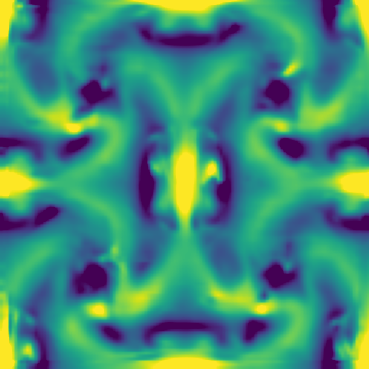}
}
\subfigure[Target DNS.]{ \label{fig:f}
\includegraphics[width=0.16\linewidth]{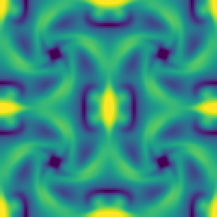}
}
\caption{Reconstructed $w$ channel by each method on a sample testing slice along the $z$ dimension in the TGV dataset. The reconstruction results are shown at 1st (80s), 5th (90s), 10th (100s), and 15th (110s) in (a)-(f), (g)-(l), (m)-(r) and (s)-(x), respectively.}
\label{fig:tf_plot4}
\end{figure*}

\subsubsection{Experimental Designs}
The proposed methods and the baselines are tested on both the FIT and the TGV datasets. The models are trained by using the FIT data from a consecutive one-second period with a time interval $\delta = 0.02s$  
and a total of $50$ time steps,  and then apply the trained model into the next $0.4$ second period (a total of $20$ time steps) for performance evaluation. For the TGV dataset, 
the models use a consecutive $40$-second period with a time interval $\delta = 2s$ for training and the next $40$ seconds of data for testing. 

The performance of DNS reconstruction is evaluated by using two different metrics, structural similarity index measure (SSIM)~\cite{wang2004image}, and dissipation ~\cite{enwiki:1127277109}. 
SSIM  is used to appraise the similarity between reconstructed data and target DNS on three aspects, luminance, contrast, and overall structure. The higher value of SSIM indicates better reconstruction performance. The dissipation operator is used to assess the performance of capturing the flow gradients. The dissipation of each of the three components of the velocity vector ($u$, $v$, and $w$) are evaluated. The dissipation operator is
defined by:

\begin{equation}
\chi (Q) \equiv \nabla Q \cdot \nabla Q= \left(\frac{\partial Q}{\partial x}\right)^2 + \left(\frac{\partial Q}{\partial y}\right)^2 + \left(\frac{\partial Q}{\partial z}\right)^2.
\end{equation}


The dissipation is used to measure the difference of  flow gradient between the true DNS and generated data. This is  represented by $|\chi({Q}^d) - \chi(\hat{{Q}}^d)|$. The lower value of this difference indicates better performance.  Compared with our previous work~\cite{bao2022physics}, the performance assessment is expanded by considering a new pixel-wise evaluation metric (dissipation) and a physical validation method based on the kinetic energy.


\subsubsection{Environmental Settings and Implementation Details}

The method is implemented via Tensorflow 2 with a GTX3080 GPU. The model is first trained in 500 epochs with ADAM optimizer~\cite{kingma2014adam_arxiv} from an initial learning rate of $0.001$. In the refinement step,  the learning rate is lowered to $0.0005$, and the training rate is iterated by 10 epochs. All the hidden variables and gating variables are in $32$ dimensions. The values of $\alpha_0$, $\alpha_1$, and $\alpha_2$ are set as  $1,\ 0.1,$ and $0.1$, respectively.

\subsection{Reconstruction Performance}
\subsubsection{Quantitative Results.}  Table~\ref{fig:table1} and Table~\ref{fig:table2} summarize the average performance over the first $10$ time steps in the testing phase on both the FIT dataset and the TGV dataset. Compared with the baselines, CNDE-based methods perform the best in both evaluations obtaining the highest SSIM value and lowest dissipation difference. 
Several observations are made: (1) When comparing the CNDE-based methods with SR baselines, DCS/MS, FNO, and FSR models, it is observed that these baseline methods cannot recover the overall flow well and get worse performance in terms of SSIM and dissipation difference. (2) Compared with the SRCNN, the CTN, which uses the LSTM model, shows a significant improvement in both evaluations. This confirms the effectiveness of a temporal model (e.g., LSTM) in capturing temporal dependency. (3) The comparison between CTN, RKTU, CNDEp-based methods, and CNDE-based methods, indicates significant improvements by incorporating each of the three components (RKTU, TEL, and refinement). In particular, the refinement method brings the most significant improvement in terms of SSIM and dissipation differences. 

\subsubsection{Temporal Analysis. } 
In the temporal analysis of the FIT dataset, the performance for reconstruction is measured for each step during a $0.4s$ period ($20$ time steps) in the testing phase. The performance change using the SSIM and the dissipation difference is shown in  Figs.~\ref{fig:tf_plot1} and ~\ref{fig:tf_plot2}, respectively. These figures indicate that: (1) With larger time intervals between training data and prediction data, the performance becomes worse. In general, the  CNDE-based methods are more stable over a long period, and show a much better performance than other methods. (2)  The temporal model (e.g. LSTM) resutls in  significant improvements in long-term predictions. (3)  The CNDE-based methods outperform the CNDEp-based methods, which demonstrate the effectiveness of test-time refinement in reducing the prediction bias in long-term prediction. (4) The  CNDEp-based methods yield a better performance after the 5th time steps compared with the temporal baseline CTN model. This indicates the advantage of the  RKTU structure in the long-term prediction.  (5) The  CNDEp-E slightly outperforms the CNDEp-R in the long-term prediction. A  similar observation is made by comparing two versions of CNDE-based methods. 

In Figs.~\ref{fig:tf_plot4_tgv} and ~\ref{fig:tf_plot5_tgv}, the results for the TGV  are presented. A better performance of the model developed here is indicated via the SSIM and dissipation differences. Several observations are made:
(1) The CNDE-based methods using refinement perform much better than CNDEp-based methods and DCS/MS. Moreover, the performance of the CNDEp-based methods becomes worse than the baseline DCS/MS after the 5th time step. This is because of the variability of TGV data over larger time intervals ($\delta= 2$s) and the testing data are very different from the initial data point.  This causes the CNDEp-based methods to fail in capturing the correct flow dynamic without refinement. It also indicates the advantages of the refinement method for adjusting the state of flow data in the long-term prediction. (2) The CTN almost fails to capture the flow dynamics after the 5th time step, thus the CTN is not suitable for this dataset.

\subsubsection{Visualization.} 
In Figs.~\ref{fig:tf_plot3},  the reconstructed data are shown at multiple  (1st, 5th, 10th, and 20th) time steps after the training period. For each time step, the slices of the $w$ component at a specified $z$ value are shown. In the 1st step, both the CNDE-based methods and the baseline CTN model yield ideal reconstruction results. This is because the test data are similar to the training data at the last time step. 
In contrast, the baseline DSC/MS~\cite{fukami2019super} leads to a poor performance starting from early time. Beginning at the 5th time step, the  CNDE-based methods perform better than the baselines. A more significant difference is observed at the 20th time step. All the baselines almost fail to capture the correct flow transport pattern. The CNDE-based methods yield a much better performance in the late stage. Similar observations are  made on the  TGV dataset as shown in Fig.~\ref{fig:tf_plot4}.

\subsubsection{Validation via  Physical Metrics.} 
\begin{figure} [!t] 
\centering
\subfigure[FIT data.]{ \label{fig:a}{}
\includegraphics[width=0.40\linewidth]{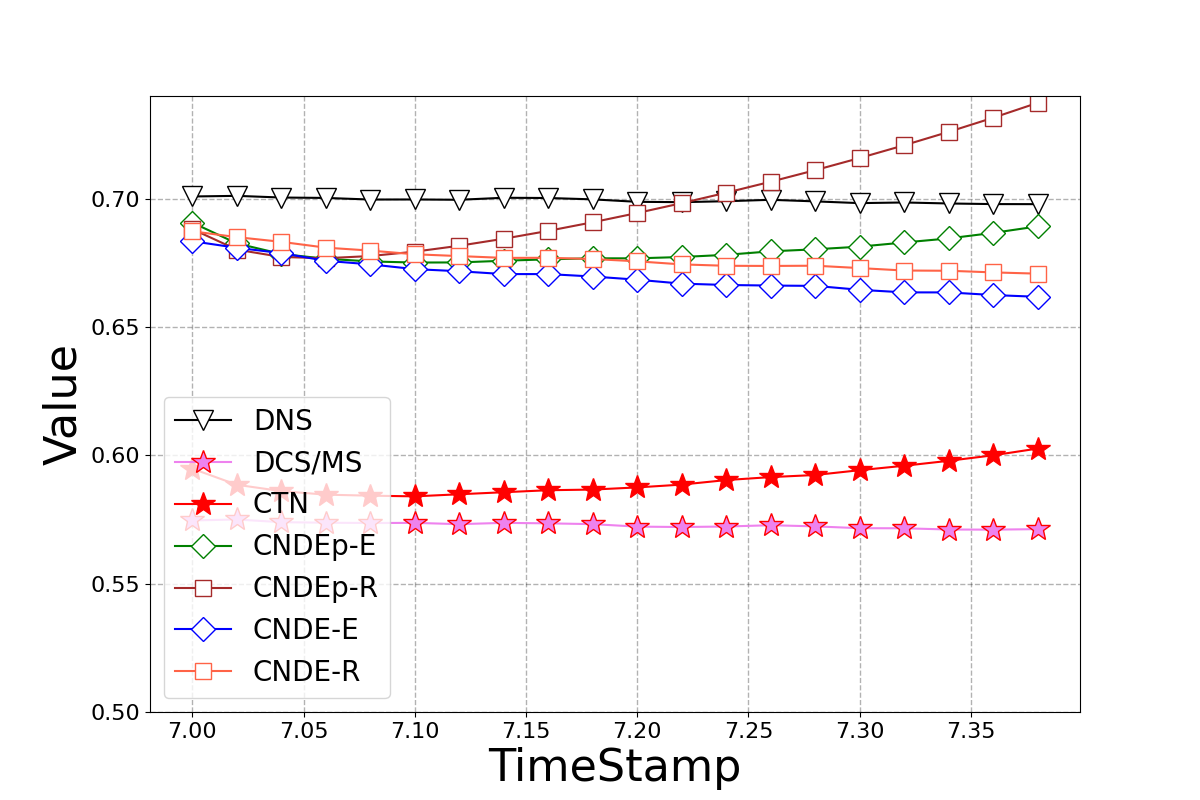}
}
\subfigure[TGV data.]{ \label{fig:b}{}
\includegraphics[width=0.40\linewidth]{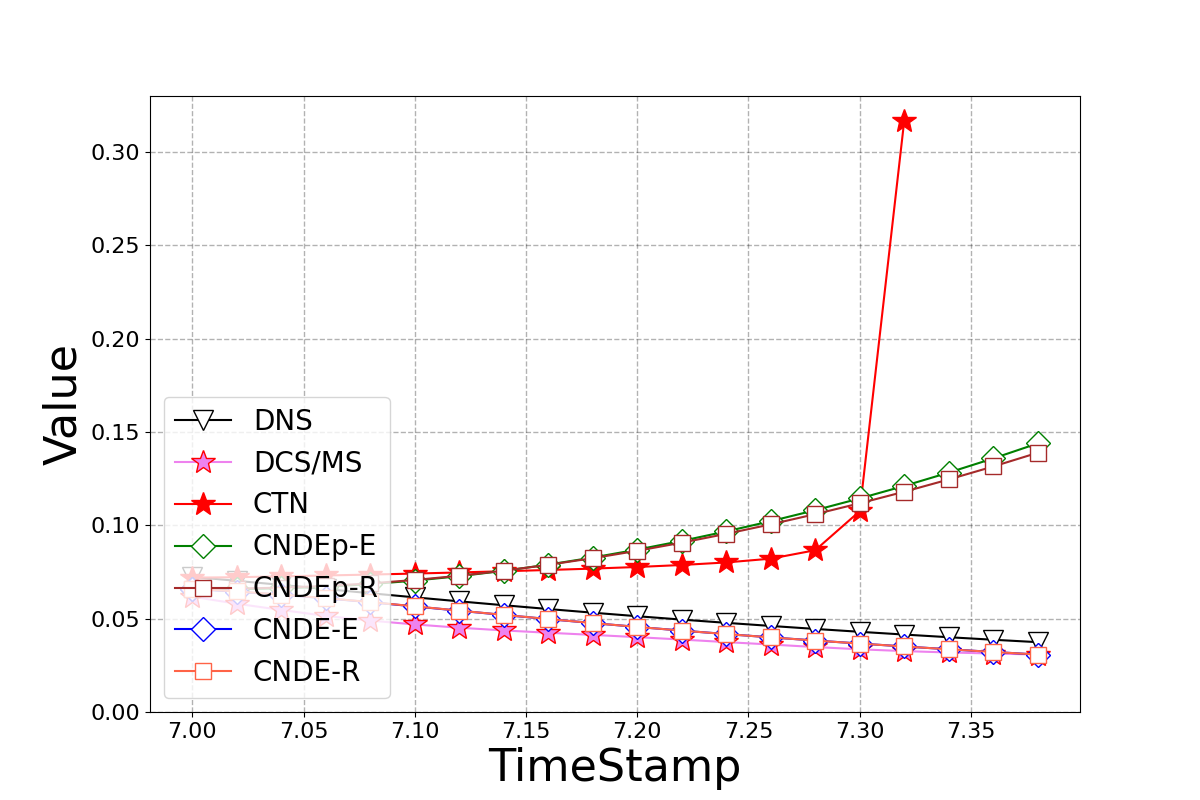}
}
\vspace{-.05in}
\caption{Change of kinetic energy produced by the reference DNS and different models in both the FIT  and the TGV datasets, respectively.}
\label{fig:kinetic}
\end{figure}

The model performance is also assessed via of long-term prediction of the turbulent kinetic energy.  Figure \ref{fig:kinetic} show the energies corresponding to the target DNS,  and the reconstructed flow data by the baselines and the  CNDE-based methods for both the FIT and the TGV flows. The results in Fig.~\ref{fig:kinetic} (a) for the  FIT dataset indicate: (1)~The CNDE-based methods in general perform better than the baseline method DCS/MS and CTN. Even without using the refinement process, the CNDEp-based methods outperform the DCS/MS and CTN models.  CNDE-based methods can follow the underlying physical rule well in the long-term prediction. 
(2) The performance of CNDEp-based methods becomes very poor after the 8th time step. This is because the accumulated error gets amplified in every time step.  The results in  Fig.~\ref{fig:kinetic} (b) yield similar conclusions.   

\section{Summary \& Concluding Remarks}

A novel super-resolution (SR) methodology, termed ``continuous networks using differential equations'' (CNDE) is developed to reconstruct high-resolution flow data in spatial and temporal fields. The model is used in the setting of unsteady,  incompressible, Newtonian turbulent flow under spatially homogeneous conditions.  The SR method is to generate the high resolution direct numerical simulation (DNS) field from low resolution, large eddy simulation (LES) data. A Runge-Kutta transition unit (RKTU) is developed to leverage the physical knowledge embodied in the Navier-Stokes equation to capture the spatial resolution, and the temporal dynamics of the flow. A temporally-enhancing layer (TEL) is constructed to capture long-term temporal dynamics. A degradation-based refinement method is developed to adjust the reconstructed data over time by enforcing the consistency with physical constraints. The performance of the model is assessed in the setting of two  flow configurations via  flow visualization and statistical analysis. The results demonstrate the superiority of the CNDE  for spatio-temporal reconstruction of the flow.   The model's constituents, the  RKTU and the refinement methods can  be used as  building blocks to enhance existing deep learning models. 

Despite its demonstrated capabilities, there are two limitations associated with the CNDE model in its current form.  (1)  The CNN layers are used to estimate spatial derivatives, which can introduce bias due to the approximation and due to data overfitting. (2) The method is, thus far,  tailored and appraised  for specific flows.   Therefore, its generality cannot be warranted for other applications; especially in the absence of sufficient DNS data. Future work is recommended to find alternative ways to evaluate the spatial derivatives more accurately and to improve the model's transferability.
 

\begin{acks}

This research is sponsored by the National Science Foundation (NSF) through   Grants OAC-2203581, IIS-2239175, and CBET-2152803. Computational resources are provided by the  University of Pittsburgh Center for Research Computing (CRC).
\end{acks}

\bibliographystyle{ACM-Reference-Format}
\bibliography{sample-base}

\begin{appendices}

\end{appendices}

\end{document}